\documentclass[12pt,epsf,amstex]{article}
\usepackage{pstricks,pst-node,amsmath,amssymb}
\usepackage[english]{babel}
\usepackage[utf8]{inputenc}
\usepackage{ae,aecompl}

\usepackage[ps2pdf,colorlinks=true,urlcolor=blue,citecolor=blue,hyperfootnotes=false]{hyperref}

\usepackage [dvips]{graphicx}
\usepackage{amsmath}
\usepackage{times}
\usepackage{epsfig}
\usepackage{amssymb}

\newcommand{\C}{\mathcal C}

\newcommand{\ZZ}{\mathcal Z}
\newcommand{\KS}{\text{KS}}

\newcommand{\dd}{\text{d}}

\newcommand{\ee}{\text{e}}

\newcommand{\p}{\partial}

\newcommand{\OO}{\text{\cal 0}}

\begin{document}
\begin{center}
{\Large{\bf Thermodynamic formalism \\
\vspace{8pt}
for systems with Markov dynamics}}
\end{center}

\vspace{2cm}
\begin{center}
{V. Lecomte$^{1,2}$, C. Appert-Rolland$^{1,3}$ and F. van Wijland$^{1,2}$}
\end{center}

\noindent ${}^1$Laboratoire de Physique Th\'eorique (CNRS UMR 8627), B\^at. 210,
Universit\'e de Paris-Sud, 91405 Orsay cedex, France.\\

\noindent ${}^2${Laboratoire Mati\`ere et Syst\`emes Complexes (CNRS
  UMR 7057), Universit\'e de Paris VII -- Denis Diderot,
  10 rue Alice Domon et L\'eonie Duquet, 75025 Paris cedex 13, France}\\

\noindent ${}^3${Laboratoire de Physique Statistique (CNRS UMR 8550), \'Ecole Normale Sup\'erieure, 24 rue Lhomond 75005 Paris, France.}\\\\

\begin{center}{\bf Abstract}\\
\end{center}
The thermodynamic formalism allows one to access the chaotic properties
of equilibrium and out-of-equilibrium systems, by deriving those from a dynamical partition function.
The definition that has been given for this partition
function within the framework of discrete time Markov chains was not suitable
for continuous time Markov dynamics. Here we propose another interpretation
of the definition that allows us to apply the thermodynamic formalism to continuous time.

We also generalize the formalism --a dynamical Gibbs ensemble construction-- to a whole family of observables and their
associated large deviation functions.
This allows us to make the connection between the thermodynamic formalism and
the observable involved in the much-studied fluctuation theorem.

We illustrate our approach on various physical systems:
random walks, exclusion processes, an Ising model and the contact process.
In the latter cases, we identify a signature of the occurrence of dynamical phase transitions.
We show that this signature can already be unraveled using the simplest dynamical ensemble one could define, based on the number of configuration changes a system has undergone over an asymptotically large time window.

\newpage

\section{Introduction}
\subsection{Motivations and outline}
In trying to bridge the microscopics of a dynamical system to its macroscopic properties, amenable to a statistical physics treatment, the main road is the study of its chaotic properties. These revolve around such concepts as Lyapunov exponents, Kolmogorov-Sinai entropy, and perhaps more refined still, that of dynamical partition function. The latter was introduced  by Ruelle (it is also called Ruelle pressure), and can be seen~\cite{ruelle} as a dynamical analog to the well-known equilibrium partition functions of statistical mechanics, except that it involves counting trajectories in phase space rather than microscopic states.
This so-called pressure, in information theoretic language, is not but the R\'enyi entropy associated with the measure over the set of possible trajectories in configuration space~\cite{beckschlogl}.
It can then be connected to the dynamical entropies, like the Kolmogorov-Sinai entropy, also viewed as the Shannon entropy over the set of trajectories, or the topological entropy, which measures the growth rate of the number of allowed trajectories.
 Back in the seventies, the dynamical partition function also appeared as a convenient tool for characterizing, under prescribed mathematical conditions, NonEquilibrium Steady-State (NESS) measures, now  called Sinai-Ruelle-Bowen (SRB) measures~\cite{ref-SRB}, by means of a variational principle. The general framework behind is that of temporal large deviations. A vast body of mathematical physics literature has been devoted to SRB measures and large deviations, with however relatively few direct spinoffs for theoretical physics, let alone experimental physics.
Actually, though these notions were mathematically well established in
various frames (Hamiltonian dynamical systems, maps, Markov chains...),
physically relevant explicit results for the Kolmogorov-Sinai entropy
are scarce, with a few exceptions for the Lorentz gas and hard-spheres~\cite{dorfmanernstjacobs,vanbeijerendorfman,vanbeijerendorfmanposchdellago}. There are also numerical studies~\cite{lienhks-diff} of simple fluids attempting to relate the Kolmogorov-Sinai entropy to the equilibrium excess entropy, or to the self-diffusion constant.
When it comes to determining the full topological pressure, existing results are confined to simple maps~\cite{dorfman} or to simple Markov processes in discrete time such
as the Lattice Lorentz Gas \cite{appertvanbeijerenernstdorfman}.
However, recent years have witnessed the revival of  large deviations, both at the experimental and  theoretical level. On the theoretical side, they appeared as the natural language in which the fluctuation theorem of Gallavotti and Cohen~\cite{gallavotticohen} was expressed. The latter can be seen as a symmetry property of the large deviation function of the entropy current resulting from driving a system into a NESS. Variations around that fluctuation relation, such as the earlier Evans-Searles~\cite{evanssearles94}, or the Jarzynski nonequilibrium work relation~\cite{jarzynski}, also rely on the concept of large deviations. The experimental motivation lies in the belief that global --{\it i.e.} space averaged-- quantities, rather than local probes, are a better way to approach and above all compare between themselves systems out of equilibrium. However, since the peculiarities of a NESS also result from its microscopic dynamics, it was suggested to measure time averaged (over a large time interval) quantities, and to build up the corresponding distribution functions. More recent experiments  on electric circuits have been used to probe the hypotheses underlying the mathematics of those relations (for a nonexhaustive list of experimental references, see \cite{ldf-exp}).

While the above deals with actual dynamical systems, there also exist Markov dynamics counterparts to many of the results mentioned above, as far as fluctuation theorems are concerned (see \cite{kurchan} or \cite{lebowitzspohn} for the fluctuation relation, and see \cite{crooks} for the nonequilibrium work relation). The motivations for addressing systems with Markov dynamics (with continuous time) are to be found both in the greater ease in performing numerical simulations (as cleverly proposed in \cite{giardinakurchanpeliti}) and in the analytical insight that can be gained through exact~\cite{derridalebowitz,derridaappert,bodineauderrida,lecomteraczvanwijland, farago} or approximate calculations~\cite{vanwijlandracz}. To the best of our knowledge, these explicit calculations have been attempted only for systems with Markov dynamics. Given the successes of the Markov approach in understanding the various versions of the fluctuation and work theorems, it seemed natural to turn to the more general dynamical partition function. As briefly sketched in \cite{lecomteappertrollandvanwijland},  by contrast with the existing treatment of Markov chains~\cite{gaspard, dorfman,dorfmanernstjacobs,appertvanbeijerenernstdorfman} there had hitherto been no satisfactory attempt to force the thermodynamic formalism of Ruelle into the framework of systems endowed with continuous-time Markov dynamics.
As this was already noticed by Gaspard \cite{gaspard2}, passing from discrete to continuous time raises specific difficulties.\\

Therefore, our primary purpose in the present paper is to introduce the dynamical partition function and the related topological (or Ruelle) pressure for systems with Markov dynamics.
Note however that our motivation for determining this dynamical partition function is not rooted in our quest for the Markov analog of an SRB measure. For finite systems with Markov dynamics this is a dull endeavor since the stationary measure is known to be the unique solution to the stationary master equation~\cite{vankampen}. Instead, we have in mind gaining physical insight into the topological pressure. It is often presented  as a measure of dynamical complexity, an interpretation which will appear quite clearly in systems with ergodicity breaking transitions. Beyond, our general goal is to be able to relate its properties (convexity, singularities, {\it etc}) to those of the system at hand, the latter displaying nontrivial dynamics, and possibly featuring strong interactions.
Ideally, we would like to build up a picture gallery~\cite{racz} for physically acceptable topological pressures, but in practice we will have to be more modest and we will focus on a restricted number of systems that we shall soon describe. Further investigations aiming at pursuing this goal, most notably for systems with glassy dynamics and for systems with quenched disorder, will be mentioned in our conclusion.\\

It will turn out that the dynamical partition function can be seen as the generating function of a physical observable. This will allow us to cast our findings into the more general framework of temporal large deviations. In setting up our mathematical approach, we will see that the latter observable is connected to --but very different from-- the one considered by Lebowitz and Spohn~\cite{lebowitzspohn}. They both are members of a rather general family of observables of which we shall further single out yet another one that we now describe.  Over a given trajectory in configuration space, the simplest quantity of all to consider is the number of configuration changes that the system undergoes over a given time interval.
While this is a seemingly trivial observable to consider, we will illustrate on specific examples that much of the difficulties that pave the way to the full determination of, say, the topological pressure, can already be read off the study of the statistics of this event-counting observable.
More important, we find that ``dynamical phase transitions'', as defined
for example in \cite{beckschlogl},
can already be observed
on this simple observable, and not only on the topological pressure.
We propose a new tool to study how the structure of the trajectory space
is affected by the dynamical phase transition.\\

We now describe the various systems that we have chosen to illustrate our approach. We begin with examining the simple lattice random walk case.
We continue with an interacting lattice gas, namely the one
dimensional exclusion process with periodic boundary conditions, for
which our analytic results are somewhat less extensive, but that has
in the recent
past~\cite{derridalebowitz,derridaappert,lebowitzspohn,derridadoucotroche}
served as a testbench for many of the ideas discussed in this
introduction.
In the case of the symmetric exclusion process we found that,
though there is no first order dynamical phase transition,
the event-counting observable
mentioned above shows signs of a second order
dynamical phase transition.
Then we turn to two mean-field models of
interacting degrees of freedom. The first one is the well-known
equilibrium Ising model, with a second order symmetry breaking phase
transition to an ordered state at low temperatures.
We have shown that the thermodynamic phase transition
induces a first order dynamical phase transition,
a signature of which can already
be found on the event-counting observable.
Besides, we were able to give a picture of the structure of the
trajectory space through the transition.
The second one is
the contact process, for which a supplementary difficulty arises,
as in the thermodynamic limit two stationary states --an active and
an absorbing one-- exist.

But before embarking into the study of
these physical systems, we devote Sec.\,II to a reminder of the
definitions of Lyapunov exponents, Kolmogorov-Sinai entropy, and also
of the state of the art~\cite{gaspard} concerning systems with
discrete-time Markov dynamics. Sec.\,III contains our construction of
the dynamical partition function for systems with continuous-time
Markov dynamics, and connects to the existing literature. Secs.\,IV,\:V
and VI are concerned with our physical examples. Conclusions and a
number of future research directions are gathered in Sec.\,VII.

\section{Kolmogorov-Sinai entropy in the theory of dynamical systems} 
\subsection{Dynamical systems}
Let $\Gamma(t)$ be the coordinate of a dynamical system evolving according
to $\frac{\dd\Gamma}{\dd t}={\cal F}(\Gamma)$. Consider now two infinitesimally
close initial points $\Gamma(0)$ and $\Gamma(0)+\delta \Gamma (0)$ and follow
the evolution of the difference $\delta \Gamma(t)$ between the two. This will
evolve according to
$\frac{\dd \delta\Gamma}{\dd t}=    \frac{\p {\cal F}}{\p\Gamma}\delta\Gamma$.
The eigenvalues of the linearized evolution operator $\frac{\p {\cal F}}{\p\Gamma}$,
once averaged with respect to the stationary measure,  make up the Lyapunov spectrum
$\{\lambda_i\}$  of the dynamical system. There are as many Lyapunov exponents
as phase space dimensions. Each of them characterizes the dynamical instability
of the system along an individual direction. A system with at least one positive
Lyapunov exponent is termed {\it chaotic}. In order to characterize global, rather
than individual, chaoticity, the Kolmogorov-Sinai entropy was defined. Given a
partition of phase space, within this coarse grained description, the dynamics
becomes probabilistic, and this allows one to construct a measure over the set
of physically realizable trajectories of the system over some time interval
$[0,t]$ (which we also call histories). We define the Kolmogorov-Sinai (KS)
entropy as the Shannon entropy corresponding to the measure over the set of histories:
\begin{equation} \label{eqn:h_KS}
 h_\KS= - \lim_{t\to\infty} \frac{1}{t} 
\frac{\sum_{\substack{\text{histories}\\ \text{from }0\to t}}
 \text{Prob}\{ \text{history} \} \ln \text{Prob}\{ \text{history} \}}{\sum_{\substack{\text{histories}\\ \text{from }0\to t}} \text{Prob}\{ \text{history} \} }
\end{equation}
where the supremum is taken over all possible partitions and the average is taken over
the initial configuration. The denominator is equal to $1$ for a close system.
From its definition, it is clear that $h_\KS$ measures the dynamical randomness of the system at hand.  It is also connected in a simple way to the Lyapunov spectrum, by means of Pesin's theorem, which states that
\begin{equation}
h_\KS=-\gamma+\sum_{\lambda_i>0}\lambda_i
\label{eqn:pesin}
\end{equation}
where $\gamma$, defined for an open system, is its escape rate (and is otherwise zero). Note that the KS entropy is defined for a system in a stationary state, in or out of equilibrium. 
Even if one would like to relate $h_\KS$, at least in equilibrium situations, to the standard Boltzmann entropy, there is no direct connection between both, the latter being an intrinsically static object while the former is dynamical in essence.
However, Boltzmann's entropy variations are related to $h_\KS$. Finally, we turn to a definition~\cite{ruelle} of the dynamical partition function $Z(s,t)$ 
\begin{equation} \label{eqn:dyn_part_func}
    Z(s,t) = \sum_{\substack{\text{histories}\\ \text{from }0\to t}}
             \big( \text{Prob}\{ \text{history} \} \big)^{1-s}
\end{equation}
In practice the so-called thermodynamic limit $t$ very large is understood. We have also substituted $1-s$ for the canonical notation $\beta$ which we keep for denoting an inverse temperature
(the reason for
introducing $s$ in this way will become obvious when we shall
express $Z$ as a generating function).
There is an alternative formulation for the dynamical partition function, which involves the local stretching factors (see {\it e.g.} \cite{vanbeijerendorfman2} for a physical example). The intensive potential $\psi_+(s)$ associated to this partition function is the topological pressure (or Ruelle pressure),
\begin{equation}\label{eqn:topo_press}
   \psi_+(s) = \lim_{t\to\infty} \frac{1}{t} \ln Z(s,t)
\end{equation}
which can also be interpreted~\cite{beckschlogl}, in information theoretic language, as the R\'enyi entropy over the set of histories. It is possible to recover $h_\KS$ from the topological pressure, $h_\KS=\psi_+'(0)$
(or $h_\KS=\psi_+'(0) - \gamma$ for an open system, with $\gamma = - \psi_+(0)$),
along with other quantities such as the topological entropy $h_\text{top}$, which measures the grows rate of the number of possible histories as time is increased, and is given by $h_\text{top}=\psi_+(1)$.
\subsection{Markov chains}
Given the definitions above, there is a natural way, as explained by Gaspard~\cite{gaspard,gaspard4}, to extend the definitions of the dynamical partition function and of the KS entropy to discrete time Markov processes. Consider a Markov process governed by the discrete-time master equation for the probability $P(\C,t)$ to be in state $\C$ after $n$ steps:
\begin{equation} \label{eqn:mastereq_discretetime}
  P(\C,t+\tau)-P(\C,t) = \sum_{\C'\neq\C}\big[ w(\C'\to\C)P(\C',t)
                                    -  w(\C\to\C')P(\C,t)  \big]
\end{equation}
where $\tau$ is the time step (and $t=n\tau$ is the elapsed time). We have denoted by $w(\C\to\C')$ the transition probability from configuration $\C$ to another configuration $\C'$. The probability of a history $\C_0\to\ldots\to\C_n$ taking place between $0$ and $t=n\tau$ reads
\begin{equation}
  P(\C_0\to\ldots\to\C_n)=P(\C_0,0) w(\C_0\to\C_1)\ldots w(\C_{n-1}\to\C_n)
\end{equation}
Note that successive configurations $\C_k$, $\C_{k+1}$ can be equal in the previous relation. The corresponding
probability of remaining in the same configuration $\C$ during a time step is
\begin{equation}
  w(\C\to\C)=1-\sum_{\C'\neq\C} w(\C\to\C')
\end{equation}
By definition of the KS entropy we may directly write that
\begin{equation} \label{eqn:h_KS_discrete}
h_\KS=-\lim_{n\to\infty} \frac{1}{n\tau} \sum_{\C_0,..\C_n}P(\C_0\to\ldots\to\C_n)\ln P(\C_0\to\ldots\to\C_n)
\end{equation}
It is easy to see \cite{gaspard2} that the above expression reduces to
\begin{align}
  h_\KS &= -\frac{1}{\tau}\sum_{\C,\C'} P_\text{st}(\C)\
           w(\C\to\C') \ln w(\C\to\C') \nonumber \\
        &= -\frac{1}{\tau}\langle\sum_{\C'}w(\C\to\C') \ln w(\C\to\C') 
                                             \rangle_\text{st}
 \label{eqn:h_KS_discrete_time} 
\end{align}
where we have introduced the stationary measure $P_\text{st}(\C)$. Several explicit calculations of this quantity can be found in Dorfman~\cite{dorfman} or in Gaspard~\cite{gaspard}.
\subsection{Taking the continuous-time limit} 
We now wish to take the continuous-time limit of (\ref{eqn:h_KS_discrete_time}). We scale the transition
 probabilities between different configurations with the time step $\tau$:
\begin{equation}
  w(\C\to\C') = \tau\, W(\C\to\C')
\end{equation}
in such a way that the master equation (\ref{eqn:mastereq_discretetime}) yields its continuous time analog when the limit $\tau\to 0$ is taken, namely
\begin{equation} \label{eqn:cont_time_Markov_eq}
  \partial_t P(\C,t) = 
    \sum_{\C'\neq\C}\big[ W(\C'\to\C)P(\C',t)
                                   -  W(\C\to\C')P(\C,t)  \big]
\end{equation}
As done in \cite{gaspard2,gaspard3}, the KS entropy defined in (\ref{eqn:h_KS_discrete_time}) can be expressed in terms of the transition rates $W$:
\begin{equation}\label{hKScontinuousnaive}
  h_\KS = -\sum_{\C,\C'} P_\text{st}(\C)\
           W(\C\to\C') \ln \left(\tau W(\C\to\C')\right) \\
\end{equation}
It is now clear that the limit $\tau\to 0$ in (\ref{hKScontinuousnaive}) does not exist, since the latter exhibits a $\ln \tau$ divergence as $\tau\to 0$. Given that the transition rates $W$ are dimensionful quantities, and given that apparently the only available time scale is $\tau$, we cannot expect to get rid of $\tau$ without further thoughts. This means that even if we were tempted to retain in (\ref{hKScontinuousnaive}) only the finite contribution as $\tau\to 0$ as the meaningful KS entropy, we would need to find an appropriate time scale to render this piece well-defined (the argument of the logarithm must be dimensionless).\\

The one-dimensional lattice random walk perhaps constitutes the simplest example of a Markov chain: let $p$ (resp. $q$, $r$) denote the probability of hopping to the right (resp. to the left, not hopping), then we have that
\begin{equation}
h_\KS=-\frac{1}{\tau}\left[p\ln p+q\ln q+r\ln r\right]
\end{equation}
It appears clearly that in the continuous-time limit, $p$ and $q$ become infinitesimally small, which produces an indefinite $h_\KS$. Since we have in mind describing as closely as possible dynamical systems, which evolve in continuous time, the goal we set ourselves is to find a consistent approach, intrinsically viable for Markov systems in continuous time.
\section{Systems with continuous-time Markov dynamics}
\subsection{Histories and dynamical partition function}
\label{subsec:histories_Z_of_s}
We now consider a system with Markov dynamics, with transition rate $W(\C\to\C')$ from configuration $\C$ to configuration $\C'$, in which the probability $P(\C,t)$ to be in state $\C$ evolves according to the following master equation,
\begin{equation}
\p_t P=\mathbb{W}P
\end{equation}
where the evolution operator has the matrix elements
\begin{equation}\label{opevolmatel}
\mathbb{W}(\C,\C')=W(\C'\to \C)-r(\C)\delta_{\C,\C'}
\end{equation}
and
\begin{equation}\label{opevolmatel2}
 r(\C)=\sum_{\C'\neq\C}W(\C\to \C')
\end{equation}
is the rate of escape from configuration $\C$.
In order to overcome the difficulties encountered in the previous section,
an alternative is to consider histories
$\C_0\to\ldots\to\C_K$ in configuration space, in the spirit of the
study by van Beijeren, Dorfman and
collaborators~\cite{vanbeijerendorfman2,latzvanbeijerendorfman} of the Lorentz gas and
the Sinai billiard.  To give (\ref{eqn:h_KS}) or (\ref{eqn:dyn_part_func}) a
consistent meaning for continuous time dynamics, we first interpret
$\text{Prob}[\text{history}]$ as the probability in the enumerable set of
histories in configuration space.  A history in configuration space is a
sequence $\C_0\to\ldots\to\C_K$ of successive configurations. An important difference between discrete and continuous time dynamics is that in the latter, the system stays in each state  $\C$ for a random time lapse drawn from an exponential distribution of parameter $r(\C)$ as defined in (\ref{opevolmatel2}), which is interpreted as the rate of escape from configuration $\C$ to any other configuration. Then the system hops to its next state $\C'$ with probability $\frac{W(\C\to\C')}{r(\C)}$. Given the initial state $\C_0$, the probability of the history $\C_0\to\ldots\to\C_K$ is the product of each jump probability
\begin{equation}
 \text{Prob}\{\text{history}\}=\prod_{n=0}^{K-1} \frac{W(\C_n\to\C_{n+1})}{r(\C_n)}
\end{equation}
where $K$ is the number of changes in configuration space.\\

We argue that in the same way as (\ref{eqn:h_KS})
and (\ref{eqn:dyn_part_func}) are averaged over the initial configuration when
dealing with deterministic dynamical systems, we similarly have to average
over all possible stochastic time sequences $t_0,\ldots,t_K$ at which
configuration changes occur ($K$ is a fluctuating number).  
We know from general properties of Markovian system~\cite{vankampen} that
the duration $\Delta
t=t_{n+1}-t_n$ between configurations $\C_n$ and $\C_{n+1}$ is distributed
according to the probability density
\begin{equation}
  \pi(\C_n,\Delta t)=r(\C_n)\ee^{-\Delta t\ r(\C_n)}
\end{equation}
\begin{figure}
\psset{unit=10mm}
\begin{center}
\begin{pspicture}(0,-.7)(11,1.2)
   \psline{|-}(0, 0)(3,0)
   \psline{|-}(3,0)(5,0)
   \psline{|-}(5,0)(9,0)
   \psline{|-|}(9,0)(11,0)
   \psline{<->}(9,0.18)(11,0.18)

   \rput[b](0,0.3){$0$}
   \rput[t](0,-.3){$\C_0$}
   \rput[b](3,0.3){$t_1$}
   \rput[t](3,-.3){$\C_1$}
   \rput[b](5,0.3){$t_2$}
   \rput[t](5,-.3){$\C_2$}
   \rput[b](7,0.3){$\ldots$}
   \rput[b](9,0.3){$t_K$}
   \rput[t](9,-.3){$\C_K$}
   \rput[b](11,0.3){$t$}
   \rput[t](11,-.3){$\C_K$}

   \rput[cb](10,1){\footnotesize {\begin{tabular}{c}
        waiting probability: \\ $ \ee^{-(t-t_K)r(\C_K)} $
      \end{tabular}}}
\end{pspicture}
\end{center}
\caption{One particular realization in time of a history
        $\C_0\to\ldots\to\C_K$  of successive configurations.
        Between $t_{k}$ and  $t_{k+1}$, the system stays in 
        configuration $\C_k$.        }
\label{fig:time_sequence}
\end{figure}
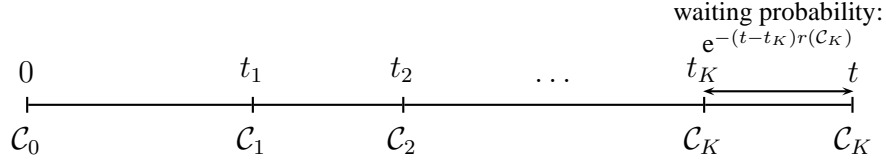
Accordingly, if we take into account every possible history
$\C_0\to\ldots\to\C_K$ between $t_0$ and $t$, the dynamical partition function
writes
\begin{align} \label{eqn:Z_markov}
 Z(s,t|\C_0,t_0) = \sum_{K=0}^{+\infty} \sum_{\C_1,\ldots,\C_K}
 & \int_{t_0}^t dt_1\: \pi(\C_0,t_1-t_0)  \ldots
   \int_{t_{K-1}}^t dt_K\: \pi(\C_{K-1},t_K-t_{K-1}) \nonumber \\ & 
 \quad
 \ee^{-(t-t_K)r(\C_K)}
 \left[
  \prod_{n=1}^K  \frac{W(\C_{n-1}\rightarrow\C_{n})}{r(\C_{n-1})} 
 \right]^{1-s}
\end{align}
where the last exponential factor $\ee^{-(t-t_K)r(\C_K)}$ is the probability not to leave state $\C_K$ in the remaining interval between $t_K$ and $t$. We have assumed (Fig.\,\ref{fig:time_sequence}) without loss of generality that
the system starts from a fixed initial configuration $\C_0$
(we restrict our study for simplicity to systems which can take
a finite number of energy states).
\subsection{Kolmogorov-Sinai entropy}
In the same spirit as for the dynamical partition function,
we interpret the definition (\ref{eqn:h_KS}) for the Kolmogorov-Sinai entropy
as
\begin{align} \label{eqn:h_KSlong}
   h_\KS = -\lim_{t\to\infty} \frac{1}{t} 
 \sum_{K=0}^{+\infty} \sum_{\C_1,\ldots,\C_K}
 & \int_{t_0}^t dt_1\: \pi(\C_0,t_1-t_0)  \ldots
   \int_{t_{K-1}}^t dt_K\: \pi(\C_{K-1},t_K-t_{K-1}) \nonumber \\ & 
 \quad
 e^{-(t-t_K)r(\C_K)}
 \left[
 \text{Prob}\{\text{history}\}
 \right]
 \ln\left[
 \text{Prob}\{\text{history}\}
 \right]
\end{align}
where we assume that the definition does not depend on the initial
configuration $\C_0$.
For simplicity, we consider only close systems (except otherwise stated).
Then we do recover the usual relation between 
 $h_\KS$ and $Z(s,t)$, i.e.
\begin{equation} \label{eqn:h_KS_long_time_limit}
   h_\KS = \lim_{t\to\infty} \frac{1}{t} 
             \left. \frac{\partial \ln Z(s,t)}{\partial s}\right|_{s=0}
\end{equation}
An a posteriori
justification of (\ref{eqn:Z_markov}) and (\ref{eqn:h_KSlong})
is that it yields a finite result, which
does not depend on any external time scale nor on a particular choice of time
units. In fact there is a natural --yet fluctuating-- time-scale $1/r(\C)$ for each state $\C$ which is occupied. Furthermore, as detailed below (see Sec.\,\ref{subsec:h_KS_R}),
the KS entropy which results from (\ref{eqn:Z_markov}) is intimately related to
the entropy flow~\cite{lebowitzspohn} of continuous time Markov processes,
exactly in the same way as was noted by Gaspard for discrete time stochastic
dynamics~\cite{gaspard2}.

We should emphasize that the definitions that we put forward in our
approach differ from those classically employed within the dynamical
systems framework. The original Kolmogorov-Sinai entropy corresponding
to the measure over histories in time and in configuration space is
infinite~\cite{gaspard}, as the information needed to describe the
continuously distributed time intervals between configuration changes
is itself infinite~\cite{grassberger}. As such, this point of view
cannot be used to compare different Markov processes in continuous
time. As explained above (Sec.~\ref{subsec:histories_Z_of_s}), we instead preferred to focus on
the information contained solely in the sequence of configurations,
handling separately the averaging over time intervals. We exemplify
below in several examples that, in doing so, the original spirit
and physical content of the Ruelle thermodynamic formalism is
preserved.

\subsection{Expressions in terms of an observable}

It is possible to express both the dynamical partition function and
 the Kolmogorov-Sinai entropy  in terms of a history dependent
observable $Q_+$ defined as
\begin{equation}\label{defQ+}
  Q_+ = \sum_{n=0}^{K-1}\ln \frac{W(\C_n\to\C_{n+1})}{r(\C_n)}
\end{equation}
We see that in the configuration space, 
\begin{equation} 
 \text{Prob}\{\text{history}\} = \ee^{ Q_+ }
\end{equation}
Hence, from (\ref{eqn:Z_markov}), $Z(s,t)$ can be identified as the generating
function of $Q_+$:
\begin{equation} \label{eqn:Z_as_generating_func}
  Z(s,t) =\langle \ee^{-s Q_+}\rangle
\end{equation}
where $\langle \cdot \rangle$ stands for an average in
both configuration and time sequences spaces.
Further using the result (\ref{eqn:Z_as_generating_func}) we also remark that
\begin{equation}
   h_\KS = - \lim_{t\to\infty} \frac{1}{t} 
             \left\langle Q_+ \right\rangle
\end{equation}
\subsection{Topological pressure}
The moment generating function of the physical observable $Q_+$ defined in (\ref{defQ+}) is precisely $Z(s,t)$. The function $\psi_+(s)$ defined by
\begin{equation} \label{eqn:psiplusdef}
\psi_+(s)=\lim_{t\to\infty}\frac{1}{t}\ln Z(s,t)
\end{equation}
is called the topological --~or Ruelle~-- pressure in analogy with (\ref{eqn:topo_press}).
This is also the generating function for the cumulants of $Q_+$:
\begin{equation}
\lim_{t\to\infty}\frac{\langle Q_+^n\rangle_c}{t}=(-1)^{n}\left.\frac{\dd^n\psi_+}{\dd s^n}\right|_{s=0}
\end{equation}
The dynamical partition function is expected to grow exponentially with time as
$\ee^{t\psi_+(s)}$, and the growth rate $\psi_+(s)$ is the topological pressure.
One immediately recognizes that the KS entropy can be obtained from $\psi_+$ through
\begin{equation}
h_\KS=\left.\frac{\dd\psi_+}{\dd s}\right|_{s=0}
\end{equation}
In order not only to mathematically justify the existence of $\psi_+(s)$, but also to relate it directly to the rates of the Markov process, we write an evolution equation for the probability $P(\C,Q_+,t)$ that the system is in state $\C$ at time $t$ with the value $Q_+$:
\begin{equation} \label{eqn:joint_mast_eqn_C_Q_+}
\begin{split}
  \partial_t P(\C,Q_+,t) = 
    \sum_{\C'\neq\C}\bigg[ & W(\C'\to\C)
          P\left(\C',Q_+-\ln \frac{W(\C'\to\C)}{r(\C')},t\right)\\
     & -  W(\C\to\C')P(\C,Q_+,t)  \bigg]
\end{split}
\end{equation}
Noticing that the average $\left\langle Q_+ \right\rangle $ 
over the configuration and time sequences is the same as
$\sum_{\C, Q_+} Q_+ P(\C,Q_+,t)$, we have
\begin{equation} \label{eqn:qplus}
\partial_t \left\langle Q_+ \right\rangle = 
\sum_{\C,\C'\neq\C}P(\C,t) W(\C\to\C')\ln \frac{W(\C\to\C')}{r(\C)}
\end{equation}
Taking the long time limit, we find that $h_\KS$ can be expressed
as the mean value in the stationary state
\begin{equation}\label{eqn:h_KS_continuous_time} 
  h_\KS = -\left\langle\sum_{\C'}
    W(\C\to\C') \ln \frac{W(\C\to\C')}{r(\C)}\right\rangle_{st}
\end{equation}
of (the opposite of) an instantaneous observable
\begin{equation}
 J_+(\C) = \sum_{\C'}
    W(\C\to\C') \ln \frac{W(\C\to\C')}{r(\C)}
\end{equation}
Compared with the definition (\ref{eqn:h_KS_discrete_time})
for discrete time, the division by $r$ allows to get rid of the time scale
inside the logarithm.

The master equation (\ref{eqn:joint_mast_eqn_C_Q_+}) also enables to have an
insight on $\psi_+(s)$. We can first point out that the Laplace transform
of the joint probability distribution function $P(\C,Q_+,t)$
\begin{equation} \label{eqn:P_C_s}
  \hat P(\C,s,t) = \sum_{Q_+} \ee^{-sQ_+} P(\C,Q_+,t)
\end{equation}
obeys the master-equation-like evolution equation
\begin{equation} \label{eqn:master_eq_Qplus}
\begin{split}
  \partial_t \hat P(\C,s,t) = \sum_{\C'\neq\C}
     \bigg[ &  W(\C'\to\C)^{1-s} r(\C')^s
              \hat P\left(\C',s,t\right)\\
            & -  W(\C\to\C')\hat P(\C,s,t)  \bigg]
\end{split}
\end{equation}
which can be written as
\begin{equation}  \label{eqn:master_eq_Qplus_short}
\p_t\hat{P}=\mathbb{W}_+\hat{P}
\end{equation}
where the evolution operator has the following matrix elements
\begin{equation} \label{eqn:ope_topopress}
\mathbb{W}_+(\C,\C')=W(\C'\to \C)^{1-s} r(\C')^s-r(\C)\delta_{\C,\C'}.
\end{equation}
Then, as
\begin{equation} \label{eqn:dynamic_part_func_average_joint_PDF}
 Z(s,t)= \sum_{\C,Q_+} \ee^{-sQ_+} P(\C,Q_+,t)
= \sum_\C \hat P(\C,s,t) 
\end{equation}
we conclude that the topological pressure
$\psi_+(s)$ is well defined by~(\ref{eqn:psiplusdef}) and appears as
the largest eigenvalue of the operator $\mathbb{W}_+$.

Likewise, in the context of deterministic dynamical system theory, the
topological pressure~$\psi_+(s)$ appears as the maximum eigenvalue of the
Perron-Frobenius operator~\cite[section~4.5]{gaspard}.
The operator (\ref{eqn:ope_topopress}) is the stochastic
counterpart to the Perron-Frobenius operator.

\subsection{Ruelle Zeta function}
The Ruelle Zeta function $\ZZ(s,z)$, as reviewed by Gaspard~\cite{gaspard}, is defined as the Laplace transform of
the dynamical partition function $Z(s,t)$ with respect to time
\begin{equation}
 \ZZ(s,z) = \int_0^\infty \dd t\, \ee^{-z t} Z(s,t)
\end{equation}
The Ruelle Zeta function is analytic in the complex variable $z$ except on
some poles. The topological pressure $\psi_+(s)$ is the pole which is the
closest to $0$, and there are systems for which $\psi_+$ is easier to access using that property. From the explicit definition~(\ref{eqn:Z_markov}) of
$Z(s,t)$ we remark that the temporal integrals are just interwoven
convolutions which factorize after Laplace transform to yield
\begin{align} \label{eqn:ZZ_markov}
 \ZZ(s,z) = \sum_{K=0}^{+\infty} \sum_{\C_1,\ldots,\C_K}
 \frac{1}{z+r(\C_K)}
 \prod_{n=1}^K
 \frac{r^{s}(\C_{n-1})\:W^{1-s}(\C_{n-1}\rightarrow\C_{n}) }
      {z+r(\C_{n-1})}
\end{align}

\subsection{Topological pressure in special cases}
\label{subsec:psi_special_case}

\subsubsection{Constant rate of escape $r$}
\label{subsubsec:rct}

From last section we remark that one situation is especially simple when
determining the topological pressure~$\psi_+(s)$: if the jump rates $W(\C\to\C')$ are uniform in
configuration space (we shall assume for definiteness that $W(\C\to\C')$
takes only two values, $0$ or $W$), the local rate of escape from the configurations
visited by the system $r(\C) = r$ becomes
independent of $\C$ and the topological pressure is Poissonian 
\begin{equation} \label{eqn:psi_plus_Poisson}
  \psi_+(s) = r \: \left[ \left(\frac{r}{W}\right)^s-1\right]
\end{equation}
This result can be seen by directly finding the largest
eigenvalue of the Perron-Frobenius operator~\eqref{eqn:ope_topopress} or by
the following line of reasoning.  In the definition~\eqref{eqn:Z_markov} for
the dynamical partition function $Z(s,t)$, the probability of configurational
histories
$\text{Prob}[\text{hist}]=\ee^{Q_+}$ depends on the history
$\C_0\to\ldots\to\C_K$ only through the number $K$ of configuration changes:
$\ee^{Q_+}=\left(\frac{W}{r}\right)^K$.  Thus, $\ee^{Q_+}$ decouples from the
average over time sequences $t_0,\ldots ,t_K$.  One can thus compute separately
the probability of this time sequence which is a convolution of exponential laws of
common parameter $r$, which all combine to yield a Poisson law:
\begin{align} 
   \int_{t_0}^t dt_1 \,r\, \ee^{-r\,(t_1-t_0)}  \ldots
   \int_{t_{K-1}}^t dt_K \,r\, \ee^{-r\,(t_K-t_{K-1})} \ \ 
  \ee^{-r\, (t-t_K)} = \ee^{-r t}\, \frac{r^K\,t^K}{K!}
\end{align}
Then $Z(s,t)$ takes the simple form (independently of the initial
configuration):
\begin{equation} \label{eqn:Z_markov_simple}
 Z(s,t) 
 \ \ = \ \
 \sum_{K=0}^{+\infty} 
 \ \ 
 \underbrace{
     \left(\frac{r}{W}\right)^K
  }_{\substack{
     \text{number of} \\ \text{histories} 
  }}
  \ \ 
  \underbrace{
     \ee^{-r t}\, \frac{r^K\,t^K}{K!}
  }_{\substack{
     \text{probability}\\ \text{of $K$ jumps }
  }}
  \ \ 
  \underbrace{
     \left[\left(\frac{W}{r}\right)^{1-s}\right]^K
  }_{\substack{
     \text{Prob}[\text{hist}]^{1-s} 
  }}
 \ \  
 =  
 \ \
 \ee^{\,t\, r \,\left[ \left(\frac{r}{W}\right)^s-1\right]}
\end{equation}
It could also have been possible to obtain this result 
by determining the Ruelle Zeta function \eqref{eqn:ZZ_markov}
\begin{align}
 \ZZ(s,z) = \frac{1}{z+r}\sum_{K=0}^{+\infty} 
 \left(\frac{W (\frac{r}{W})^{1+s}}{z+r}\right)^K
= \frac{1}{z-r\, ( (\frac{r}{W})^{s}-1)} = \frac{1}{z-\psi_+(s)}
\end{align}

The computation was greatly simplified because all jumps of the history are
identical and independent.

\subsubsection{Random walk with reflecting boundary conditions}

To see what happens when jumps are not identical,
we can consider a particle jumping on a chain of three sites with reflecting
boundary conditions. All jumps occur a the same rate $1$ except for one, whose
rate is $w$. The corresponding Markov matrix and the Perron-Frobenius operator
read
\begin{equation}
  \mathbb W = 
  \begin{pmatrix}
    -1 & 1    &  0 \\
     w & -w-1 &  1 \\
    0  & 1    & -1
  \end{pmatrix} 
  \qquad \text{,} \qquad
  \mathbb W_+ = 
  \begin{pmatrix}
    -1       & (w+1)^s    &  0 \\
     w^{1-s} & -w-1       &  1 \\
    0        & (w+1)^s    & -1
  \end{pmatrix} 
\end{equation}
The topological pressure follows
\begin{equation}
  \psi_+(s) = \frac{1}{2} \left\{
   -2-w+ w^{-\frac{s}{2}} 
   \sqrt{w^{s+2}+4(1+w)^s (w^s+w)}
  \right \}
\end{equation}
and it does  not correspond to $Q_+$ being Poissonian.
  
\subsection{Large deviation formalism, time-reversed KS entropy, and entropy flow}
\label{subsec:h_KS_R}
As explained in Gaspard~\cite[section~4.2]{gaspard}, a variety of dynamical ensembles can be constructed following a similar procedure as the one we followed with the variable $Q_+$. In fact, any time integrated observable $A(t)$ constructed with an arbitrary function $\alpha$ according to
\begin{equation}\label{defA}
A(t)=\sum_{n=0}^{K-1}\alpha(\C_n,\C_{n+1})
\end{equation}
with $K$ the number of configuration changes undergone by the process over the time interval $[0,t]$, can be exploited in the same vein. Admittedly, a limited number of choices will be physically relevant.

Due to the specific form of $A$, a master equation can be written for $P(\C,A,t)$, and the Laplace transform $\hat{P}_A(\C,s,t)=\sum_A\ee^{-s A}P(\C,A,t)$ will then evolve according to $\p_t \hat {P}_A=\mathbb{W}_A\hat{P}_A$, where
\begin{equation}
\label{eqn:operatorA}
\mathbb{W}_A(\C,\C')=W(\C'\to\C)\ee^{-s\alpha(\C',\C)}-r(\C)\delta_{\C,\C'}
\end{equation}
The largest eigenvalue $\psi_A(s)$ of $\mathbb{W}_A$, with eigenvector
 $\tilde{P}_A(\C,s)$, is the generating function of the cumulants of $A$,
 $\psi_A(s)=\lim_{t\to\infty}\frac{1}{t}\ln\langle \ee^{-s A}\rangle$.
 This is a convex function of $s$. One can also access the first moment of $A$
 in the long time limit, $\lim_{t\to\infty} \langle A\rangle/t=\langle J_A(\C)\rangle_{st}$, with the related current $J_A(\C)=\sum_{\C'}W(\C\to\C')\alpha(\C,\C')$, relying on the sole knowledge of the stationary state distribution.  Besides, given that there exists a master equation governing the evolution of $P(\C,A,t)$, its positivity is conserved, which means that also $\hat{P}_A(\C,s,t)\geq 0$ at all times, and consequently
\begin{equation}\label{vectpropreA}
\lim_{t\to\infty}\ee^{-\psi_A(s) t}\hat{P}_A(\C,s,t)=\tilde{P}_A(\C,s)
\end{equation}
is also positive. This allows, after appropriate normalization, to interpret
the eigenvector $\tilde{P}_A$ as a probability distribution. Direct numerical access to $\psi_A(s)$ but also to $\tilde{P}_A(\C,s)$, as recently proposed  in \cite{giardinakurchanpeliti}, can be achieved by constructing an auxiliary Markov process (based on $\mathbb{W}_A$) whose stationary distribution is precisely the normalized $\tilde{P}_A$. Much physical insight can be gained from $\tilde{P}_A$, as we shall see throughout the study of the Ising model and the contact process.\\

An interesting $A$ variable that we will spend quite some time on is the one obtained by setting in (\ref{defA}) $\alpha=1$: this is $K(t)$, the number of configuration changes that have occurred over $[0,t]$. This is certainly the simplest one to consider, which does not make its properties any trivial at all
(at first sight one would be tempted to see $K$ as Poisson distributed, which is wrong in most cases).
It will further be shown to be intimately connected to $Q_+$. We postpone our discussion to the treatment of our physical examples.\\ 

Another prominent variable is the {\it action functional} introduced by Lebowitz and Spohn~\cite{lebowitzspohn}, endowed with the meaning of an integrated entropy flow, defined by
\begin{equation}
Q_S=\sum_{n=0}^{K-1}\ln\frac{W(\C_n\to\C_{n+1})}{W(\C_{n+1}\to \C_n)}
\end{equation}
This is the observable whose cumulant generating function
 $\psi_S(s)=\lim_{t\to\infty}\frac{1}{t}\ln\langle\ee^{-s Q_S}\rangle$
 verifies the symmetry property 
 $\psi_S(s)=\psi_S(1-s)$, which is one of the possible formulations of the well-known
fluctuation theorem~\cite{evanscohenmoriss,evanssearles02,gallavotticohen,gaspard,kurchan,lebowitzspohn,maes_fluctuationrelation}.
 In boundary or field driven systems~\cite{lebowitzspohn, bodineauderrida, bertinidesolegabriellijonalasiniolandim2,bertinidesolegabriellijonalasiniolandim3, bertinidesolegabriellijonalasiniolandim4,lecomteraczvanwijland},
 for instance, this entropy flow is simply proportional to the particle current
 flowing through the system, the proportionality constant being the force
 driving the system out of equilibrium (a chemical potential or a temperature
 gradient, an applied field, {\it etc.}). It is characterized by a nontrivial
 large deviation function only for nonequilibrium systems (more precisely
those breaking
 detailed balance but for which $W(\C\to\C')\neq 0$ only if  $W(\C'\to\C)\neq 0$).
 In general, this entropy flow is a linear combination of the various currents
 (charge, particles, energy, momentum) forced by an external drive, weighted with
 the conjugate forces (or affinities). The interpretation of $Q_S$ as
 an integrated entropy flow follows from the
 remark~\cite{lebowitzspohn,gaspard3,maes_fluctuationrelation} that the
 time-dependent entropy $S(t)=-\sum_\C P(\C,t)\ln P(\C,t)$ evolves according to
\begin{equation}\label{entropyevol}
\frac{\dd S}{\dd t}=\sigma_\text{irr}+\sigma_\text{f}
\end{equation}
where $\sigma_{\text{irr}}$ is defined by
\begin{equation}
\sigma_{\text{irr}} = \frac{1}{2} \sum_{\C,\C'} \left[
W(\C\to\C') P(\C,t) - W(\C'\to\C) P(\C',t)\right]
\ln \frac{P(\C,t)W(\C\to\C')}{P(\C',t)W(\C'\to\C)}
\end{equation}
and 
verifies $\sigma_{\text{irr}}\geq 0$, with equality iff the system reaches equilibrium (with detailed balance $P_\text{eq}(\C')W(\C'\to\C) = P_\text{eq}(\C)W(\C\to\C')$).
The second term $\sigma_\text{f}$ is the entropy flow: it arises
from the external forces that drive the system into a nonequilibrium
steady-state, for which $\sigma_\text{f}=-\sigma_\text{irr}\leq 0$
and
\begin{equation}\label{entropyevol2}
\sigma_\text{f}=-\langle J_S(\C)\rangle_{st}=-\lim_{t\to\infty}\frac{\langle Q_S\rangle}{t}
\end{equation}
where
$J_S(\C)=\sum_{\C'}W(\C\to\C')\ln\frac{W(\C\to\C')}{W(\C'\to \C)}$.

It is of course desirable to make contact between entropy and the entropy variation rates $\sigma_\text{irr}$ or $\sigma_\text{f}$ and the dynamical entropies. In order to achieve that goal, we dwell into  the presentation of Gaspard~\cite{gaspard2} (carried out for discrete time evolution) by introducing an additional observable $Q_-$ describing time-reversed trajectories,
\begin{equation}\label{defQ_-}
Q_-(t)=\sum_{n=0}^{K-1}\ln\frac{W(\C_{n+1}\to\C_n)}{r(\C_{n+1})}+\ln\frac{r(\C_K)}{r(\C_0)}
\end{equation}
The additional piece  $\ln\frac{r(\C_K)}{r(\C_0)}$ appearing in (\ref{defQ_-}) stands for aesthetic reasons: it is non-extensive in time and could have been dropped without any physical consequence. There exists a corresponding cumulant generating function $\psi_-(s)$ and related time-reverse KS entropy $h_\KS^{R}$:
\begin{equation}
h_\KS^{R}=\left.\frac{\dd\psi_-}{\dd s}\right|_{s=0}=-\lim_{t\to\infty}\frac{\langle Q_-\rangle}{t}=-\langle J_-(\C)\rangle_{st}
\end{equation}
with $J_-(\C)=\sum_{\C'}W(\C\to\C')\ln\frac{W(\C'\to\C)}{r(\C)}$. By construction one immediately notices that
\begin{equation}\label{connexion}
\text{(i) }Q_S=Q_+-Q_-,\;\;\;\;\;\;\;\text{(ii) }J_S=J_+-J_-
\end{equation}
and, in the steady state,
\begin{equation}\label{connexion2}
\text{(iii) }\sigma_\text{f}=h_\KS-h_\KS^{R}
\end{equation}
Equality (iii) in (\ref{connexion2}) also appears in the dynamical system literature: $h_\KS$ (resp. $-h_\KS^{R}$) is the sum of the positive (resp. negative) Lyapunov exponents and therefore $\sigma_\text{f}$ is indeed the phase-space contraction rate (the sum of all Lyapunov exponents). We have of course no such a microscopic interpretation within the Markovian framework. Note that equalities (i) and (ii) in (\ref{connexion}) hold for fluctuating variables. 

\subsection{Analyticity breaking of the large deviation functions}

  In general, for small $s$, $\psi_A(s)$ comes as the
  eigenvalue of a perturbation of the (unique) stationary state. The uniqueness
  implies in particular that this function is analytic in the vicinity
  of~$0$.  However, it can happen that for $s$ larger than some threshold
  value $s_c$, it has to be obtained from the perturbation of a state
  which is not the stationary state anymore
(we notice that since $\mathbb W_A(s)$ is not a stochastic
matrix anymore for $s\neq 0$,
the Perron-Frobenius theorem does
  not apply and the maximal eigenvalue of $\mathbb W_A(s)$ can cross
  another eigenvalue while varying $s$).
  In that case, we may have to examine the whole spectrum to determine~$\psi_A(s)$
  for $s>s_c$.

Then, $\psi_A(s)$ need not be analytic on
  the whole real line. Such ``dynamical phase transitions''
have already abundantly been studied 
in the case of $\psi_+(s)$ \cite{beckschlogl}.
But such ``dynamical phase transition'' can also be observed
for other observables. An example in the case of $\psi_K(s)$ is
given in Sec.\,\ref{subsec:Ising_spin_system}.

In some cases the situation is even worse: it happens that systems present two
  stationary states in the thermodynamic limit, an \emph{absorbing} state and
  a {non-trivial} one
(when the number of degrees of freedom becomes infinite,
the Perron-Frobenius theorem does not apply either).
In that case, the change of perturbed state can
  occur precisely at $s=0$, and $\psi_A(s)$ may not be analytic at $s=0$. 
  An example of such a situation  is studied in
  Sec.\,\ref{subsec:contact_process}.

\subsection{State-dependent dynamical entropies $h_\KS[P]$, $h^R_\KS[P]$ }
The Kolmogorov-Sinai entropy is intended to represent the dynamical randomness
of a system when following its evolution in phase space. When a
system evolves in time starting from an initial state~$P$ which is not the
stationary solution to the master equation, we expect the dynamical randomness
to evolve in time, or in other words, to depend on the state of the system. Expression (\ref{eqn:h_KS_continuous_time}) strongly suggests to introduce the
\emph{state-dependent} dynamical entropies $h_\KS[P]$, $h^R_\KS[P]$ through
\begin{align}\label{eqn:h_KS_P} 
  h_\KS[P] &= -\langle\sum_{\C'}
              W(\C\to\C') \ln \frac{W(\C\to\C')}{r(\C)}\rangle_P \\[2mm]
           &= -\sum_{\C,\C'} P(\C)
              W(\C\to\C') \ln \frac{W(\C\to\C')}{r(\C)} 
\end{align}
and similarly
\begin{align}\label{eqn:h_KS_R_P} 
h_\KS^R[P] &= -\sum_{\C,\C'} P(\C)
              W(\C\to\C') \ln \frac{W(\C'\to\C)}{r(\C)} 
\end{align}
We study the example of an infinite range Ising spin system in
Sec.\,\ref{subsubsec:h_KS_P_spins}

\section{Physical example 1: Random walks}

This simple example provides an interesting illustration of the difference
between discrete and continuous time dynamics (Sec.\,\ref{subsec:RW}).
When the particle moves on a lattice with open boundaries, it also constitutes
an example of a system with escape (Sec.\,\ref{subsec:RW_with_escape}).

\subsection{Single random walk on a lattice}
\label{subsec:RW}

\subsubsection{Discrete time random walk}

We consider a particle moving on an infinite $d$-dimensional square lattice.
It hops with probability $D \tau$ from one site to its $2d$ neighbors at each
time step of duration $\tau$ of its evolution. The probability of not moving
at each time step is $1-2dD\tau$.
The stationary state is
uniform.
The probability of a history  of $n=t/\tau$ steps with $m$ particle jumps 
is equal to $(D\tau)^m (1-2dD\tau)^{n-m}$. The dynamical partition function
writes 
\begin{align}
  Z(s,n\tau) &= \sum_{m=0}^n \binom{n}{m} (2d)^m 
 \left[(D\tau)^m (1-2dD\tau)^{n-m}\right]^{1-s} \\
             &= \left[2d(D\tau)^{1-s}+(1-2dD\tau)^{1-s}\right]^n
\end{align}
The topological pressure is
\begin{equation}
 \psi_+(s)=\frac{1}{\tau} \ln \left[2d(D\tau)^{1-s}+(1-2dD\tau)^{1-s}\right]
\end{equation}
and the KS~entropy
\begin{equation}
  h_\KS=-2dD\ln D\tau - \frac{1}{\tau}(1-2dD\tau)\ln (1-2dD\tau)
\end{equation}
When the time step $\tau$ is adjusted so that the particle moves at each
time step ($2dD\tau = 1$), we simply find
\begin{equation} \label{eqn:psi_h_KS_ct_RW}
 \psi_+(s)= 2 d D s \ln 2d 
 \qquad\text{ and }\qquad
 h_\KS = \psi_+'(0) = 2 d D  \ln 2d 
\end{equation}
When sending the time step $\tau$ to zero, we have
\begin{equation}
 \psi_+(s)= -2 d D (1-s) +2d D^{1-s} \tau^{-s} + {\cal O}(\tau)
\end{equation}
and
\begin{equation}
 h_\KS = 2 d D  (1-\ln D\tau) + {\cal O}(\tau)
\label{hks_rw_discrete}
\end{equation}
As seen in the general case, the limit $\tau=0$ is not well defined.

\subsubsection{Continuous time random walk}

\label{sect:continuousRW}

We consider the continuous time version of the random walk considered in the
previous section: the particle now jumps with rate $W(\C\rightarrow\C')=D$
to one of its
neighboring sites.  The topological pressure can be obtained directly from
the expressions (\ref{eqn:psi_plus_Poisson})-(\ref{eqn:Z_markov_simple})
for a constant rate of escape 
\begin{align}
 Z(s,t) = \ee^{\,t\, 2dD \,\left[ \left(2d\right)^s-1\right]} ;
 \;\;\;\;\;\;\psi_+(s) = 2dD\, ( (2d)^{s}-1)
\end{align}
and
\begin{align}
 h_\text{KS} = \psi_+'(0) &=  2dD \ln 2d \label{hks_rw_continuous}\\
 h_\text{top} = \psi_+(1)  &=  2dD\:(2d-1) 
\end{align}
We remark from~(\ref{eqn:psi_h_KS_ct_RW}) that even if the KS~entropy is the
same as in the discrete time random walk with time step $\tau=1/(2d D)$, the
two topological pressures differ. The fact that both KS~entropies have the same
expression is a simple consequence of the
relations~(\ref{eqn:h_KS_discrete_time}) and~(\ref{eqn:h_KS_continuous_time})
and from the observation that in the continuous time RW, the rates of escape
$r(\C)$ do not depend on the position of the particle. 
Then, the discrete and continuous time dynamics are simply
related by choosing the discrete time step $\tau$ to be equal to the inverse of
the configuration-independent rate of escape $r(\C)=r$.

On the contrary, we can interpret relation~(\ref{eqn:h_KS_continuous_time}) by
saying that in the continuous time approach, the relevant time step $\tau$ differs upon
each jump and is equal to the inverse of the configuration-dependent 
rate of escape $r(\C)$.

In any case, one should keep in mind that, though 
(\ref{hks_rw_discrete}) and (\ref{hks_rw_continuous})
give the same expressions, they were obtained for different
definitions of $h_\KS$.

It can also be noted that, if one defines a Lyapunov exponent for
the random walk through an equivalent one-dimensional map, as described
in \cite{ernstdorfman,dorfman,lecomteappertrollandvanwijland},
we recover Pesin's theorem (\ref{eqn:pesin}).

\subsection{Random walk with open boundaries: an example of system 
 with escape}
\label{subsec:RW_with_escape}

Consider now a $d$-dimensional lattice, infinite in $d-1$ directions and
finite of width~$\ell$ in the remaining direction, embedded with absorbing
boundaries. The Perron-Frobenius operator $\mathbb W_+ $ splits in a direct
sum of~$d$ one-dimensional operators corresponding to the $d$ independent
directions of the lattice
\begin{equation}
 \mathbb W_+ = \mathbb W_+^{(\ell)}  \oplus 
               \mathbb W_+^{(\infty)}\oplus\ldots\oplus
               \mathbb W_+^{(\infty)}
\end{equation}
with
\begin{equation}
   \mathbb W_+^{(\ell)} =
   \underbrace{
   \begin{pmatrix}
            -2 D & D (2d)^s & & & (0)       \\
        D (2d)^s & -2 D     & D (2d)^s &  & \\
   &    \ddots   & \ddots   & \ddots   &    \\
   &&   D (2d)^s & -2 D     & D (2d)^s       \\
(0)&&&  D (2d)^s & -2 D     &   
\end{pmatrix}}_{\ell \text{ elements}}
\end{equation}
and $\mathbb W_+^{(\infty)}$ is the infinite version of 
$\mathbb W_+^{(\ell)}$. The topological pressure $\psi_+(s)$ is the maximum
eigenvalue of $\mathbb W_+$. We find
\begin{equation}
 \psi_+(s) = 2D  \left[ (2d)^s \left(\cos \frac{\pi}{\ell+1} -1 \right)
                      +d \big((2d)^s-1 \big)      \right]
\end{equation}
from which we get the escape rate
\begin{equation}
 \gamma = - \psi_+(0) = 2 D \left(1- \cos \frac{\pi}{\ell+1}\right)
\end{equation}
expanding for large $\ell$ to
\begin{equation} \label{eq:transport_escape}
 \gamma =   D \frac{\pi^2}{\ell^2}
\end{equation}
Gaspard and Nicolis~\cite{gaspardnicolis90} have shown that such relation holds
in the discrete time approach. Our continuous-time approach preserves the
link~\eqref{eq:transport_escape} between transport properties and escape rate in open systems.

\subsection{Many particle random walk: different points of view for $h_\KS$}\label{Extensivity}

We now consider $N$ independent random walkers on a lattice of $L$ sites
with periodic boundary conditions.
Each one still hops with rate $D$, so that $r(\C) = 2dND$.
Then, with the same calculation as in Sec.\,\ref{sect:continuousRW},
we find
\begin{equation}
 \psi_+(s) = 2dND\, \left[ (2dN)^{s}-1\right]
\end{equation}
The topological pressure, and the KS entropy $h_\KS = \psi_+'(0) =
2dND\, \ln(2dN)$, are not extensive in $N$ anymore.
It could have been tempting, as the particles are independent,
to rather write $Z_N(s,t) = (Z_1(s,t))^N$, and then the
topological pressure $2dDN\left[ (2d)^s-1\right]$ would have been
extensive. The difference comes from the fact that in the
first case, the order in which particles jump has been considered
as part of the configurational trajectory, and not in the last
case.
The first approach seems the correct one, as it can be generalized
to interacting particles. Besides, as we shall see in the next section,
the non-extensivity of $h_\KS$ was already present in discrete time
with sequential update and is thus not specific to the continuous time
limit.

\section{Physical example 2: Exclusion processes}

We now consider interacting particles, more precisely,
a simple exclusion process, i.e. a gas of $N$ mutually excluding
particles diffusing on a one-dimensional periodic lattice of $L$ sites. We denote a generic configuration of the system by $\boldsymbol
n=(n_1,\ldots,n_L)$, with $n_i=1$ when site $i$ is occupied by a particle or
$n_i=0$ otherwise.

\subsection{Totally Asymmetric Simple Exclusion Process (TASEP)}

We first consider the Totally Asymmetric Simple Exclusion Process (TASEP) where
particles can only jump to the site on their right.
Though the full calculation of the topological pressure is 
quite intricate, $h_\KS$ is much simpler to obtain via
its expressions \eqref{eqn:h_KS_discrete_time} or \eqref{eqn:h_KS_continuous_time}.
We calculate it now for various types of dynamics.

\subsubsection{TASEP: parallel updating}

At each time step $\tau = 1$, each particle may go forward with probability $p$
if the site in front is empty.

Let $n_c$ be the number of clusters in a configuration $\C$.
There are $\binom{n_c}{k}$ configurations $\C'$ which are obtained
from $\C$ by moving $k$ particles. The corresponding transition
probability is $w(\C\rightarrow \C') = p^k (1-p)^{(n_c-k)}$.
Then
\begin{equation}
  h_\KS = -\langle n_c \rangle_{st} \left[ p \ln p + (1-p)\ln (1-p) \right]
\simeq -L \rho(1-\rho) \left[ p \ln p + (1-p)\ln (1-p) \right]
\end{equation}
for large systems.

\subsubsection{TASEP: random sequential updating}

\label{sec:taseprand}

At each time step $\tau = 1/L$, one bond $(i,i+1)$ is chosen randomly. If $n_i(1-n_{i+1}) = 1$,
the particle jumps forward with probability $p$.

If a configuration $\C'$ can be obtained from $\C$ by moving exactly $1$ particle,
the corresponding transition probability $w(\C\rightarrow \C') = \frac{p}{L}$.
There are $n_c$ such configurations $\C'$.
The probability to stay in the same configuration is $w(\C\rightarrow \C) = 1-n_c\frac{p}{L}$.
To leading order in $L$ we find
\begin{equation}
  h_\KS = p \rho(1-\rho) L \ln L + \OO(L)
\end{equation}
Thus $h_\KS$ is non extensive though the dynamics is still discrete in time
(and thus though $h_\KS$ is still defined using (\ref{eqn:h_KS_discrete_time})).
 
\subsubsection{TASEP: continuous time dynamics}

For the continuous time dynamics, the transition rate between configuration
$\boldsymbol n$ and $\boldsymbol n'=(\ldots,1-n_i,1-n_{i+1},\ldots)$
is $W(\boldsymbol{n} \rightarrow \boldsymbol{n'}) = D n_i (1-n_{i+1})$.

In order to find a finite value for $h_\KS$, we are now using the definition
\eqref{eqn:h_KS_continuous_time}. We note that,
at fixed number of particles $N=\sum_i n_i$, the stationary state is uniform
and each configuration has probability
$
  P_\text{st}(\boldsymbol{n}) = 1/\binom{L}{N}
$.
Besides, $W \ln W$ is equal to $W \ln D$. Thus the KS entropy can be written
\begin{eqnarray} \label{eqn:h_KS_SEP_before_expansion}
  h_\KS & = & \: \big\langle r(\boldsymbol n) \ln \frac{r(\boldsymbol n)}{D} \big\rangle_{st}
\end{eqnarray}
where the instantaneous rate of escape $r(\boldsymbol n)= D \sum_i n_i (1-n_{i+1})$.
As the steady state is perfectly random, we see~\cite{derrida98} that all
$k$-point correlation functions $C_k$ have simple expressions:
\begin{eqnarray}\label{eqn:SEP_k-point_correl}
  C_1 &\equiv &\langle n_1 \rangle_{st}    = \frac{N}{L} \\
  C_2 &\equiv &\langle n_1n_2 \rangle_{st} = \frac{N(N-1)}{L(L-1)} \\
  C_M &\equiv &\langle n_1n_2\cdots n_M \rangle_{st} = \frac{N(N-1)\cdots (N-M+1)}{L(L-1)\cdots (L-M+1)} 
\end{eqnarray}
In the thermodynamic limit $N\to \infty$, $L\to \infty$ with $N/L=\rho$,
we get
$
\frac{\langle r(\boldsymbol n) \rangle_{st}}{L} \to D\rho(1-\rho)
$.
For finite systems, the mean value of the instantaneous rate of escape $r(\boldsymbol n)$
is, taking finite size corrections into account,
\begin{equation}
  \langle r(\boldsymbol n)\rangle_{st} = DL 
   \left(
     \frac N L   - \frac {N(N-1)}{L(L-1)} 
   \right)
\end{equation}
Then $r(\boldsymbol n)$ can be split into two parts, its
mean value, of order $L$, and a fluctuating part defined as
\begin{equation}
 r(\boldsymbol n)= \langle r(\boldsymbol n)\rangle_{st}(1+\xi / \sqrt L)
\qquad \text{ from which we get } \qquad
 \xi = \sqrt L\: \frac{r(\boldsymbol n)-\langle r(\boldsymbol n)\rangle_{st}}{\langle r(\boldsymbol n)\rangle_{st}}
\end{equation}
By definition, to all orders in $L$, we have $\langle \xi \rangle_{st} = 0$.
Moreover,
\begin{equation}
  \langle\xi^2\rangle_{st}=L\frac{\langle r(\boldsymbol n)^2\rangle_{st}-\langle r(\boldsymbol n)\rangle_{st}^ 2}{\langle r(\boldsymbol n)\rangle_{st}^ 2}
\end{equation}
Once the expression for $\langle r(\boldsymbol n)^2\rangle_{st} = 4 L\Big[ 
  C_1 - C_2
  +
  (L-3)\big(C_2-2C_3+C_4 \big)
\Big]  $ is obtained,
we get the exact expression
\begin{equation}
  \langle \xi^2 \rangle_{st} = 
\frac
  {L(N-1)(L-N-1)}
  {N(L-N)(L-2)}
\end{equation}
which expands in powers of $L$ as $\langle \xi^2 \rangle_{st} = 1 + \mathcal O(1/L)$.
This allows us to expand $h_\KS = \langle r(\boldsymbol n) \ln \frac{r(\boldsymbol n)}{D} \rangle_{st}$
through
\begin{equation}
  h_\text{KS} = \langle r(\boldsymbol n) \rangle_{st} \ln \langle \frac{r(\boldsymbol n)}{D}
 \rangle_{st} + 
   \frac{1}{2}  \langle r(\boldsymbol n) \rangle_{st}
   \frac{\langle \xi^2\rangle_{st} }{L} +
   \mathcal O (1 / L)
\end{equation}
Denoting $\sigma=D \rho(1-\rho)$
and collecting all terms, we find
\begin{equation}
  h_\text{KS} = 
  L \sigma \ln (L \sigma) + \sigma \ln (L \sigma) 
  + \frac 3 2  \sigma + \frac{\sigma}{L}\ln L + \mathcal O (1 / L)
\label{eqn:hks_tasep}
\end{equation}
We could also have developed $r$ around its thermodynamic limit 
$\langle r(\boldsymbol n) \rangle_{st}/L \to D\rho(1-\rho)$.
Then $\langle \xi \rangle_{st} \neq 0$ but
$\langle \xi^2 \rangle_{st} = 1 + \mathcal O(1/L)$.

For the TASEP model, the number of configuration changes $K$ is equal
to the total distance covered by all the particles within a certain time
interval.
The large deviation function associated to this quantity has already
been calculated both for in the large system
size limit and for finite systems in \cite{derridalebowitz,derridaappert}.

\subsection{Symmetric Exclusion Process (SEP)}

We now consider the Symmetric Exclusion Process (SEP) where each
particle hops with equal probability per unit time $D$ to its right or left --~if the
target sites are empty.

In this case we have calculated not only the KS entropy but also
the large deviation function associated with the observable $K(t)$.
Though this is a simpler observable than $Q_+$, the complexity of
the calculations is already present. It gives a cruder physical picture of the level of activity undergone by the system's dynamics than $Q_+$.

\subsubsection{The Kolmogorov-Sinai entropy}

The same expression (\ref{eqn:hks_tasep})
as for TASEP holds, but now with $\sigma=2 D \rho(1-\rho)$. For the SEP, the compressibility and the strength  of the equilibrium current fluctuations, as defined by Bodineau and Derrida~\cite{bodineauderrida}, are closely intertwined\footnote{The coefficients $D(\rho)$ and $\sigma(\rho)$ appearing in \cite{bodineauderrida} verify $\sigma(\rho)/D(\rho)=2\rho^2 k_\text{\tiny B} T\chi(\rho)$, where $\chi$ is the thermodynamic isothermal compressibility and $T$ is the equilibrium temperature. For the SEP one has $D(\rho)=D$ and $\sigma(\rho)=2D\rho(1-\rho)$.} \cite{vanwijlandracz}.
Thus
one may speculate whether for another equilibrium model of interacting particles, and beyond, for a realistic interacting gas, $h_\text{KS}$ can be expressed solely in terms of the thermodynamic compressibility. This issue, which is reminiscent of earlier discussions~\cite{lienhks-diff} certainly deserves further investigation.

\subsubsection{Number of hops}
Let $K(t)$ be the number of hops performed by the Markov process on $[0,t]$ and
let $P(K,t)$ be the probability distribution function of $K(t)$. We also introduce the moment generating function
$\hat{P}_K(s,t)$ defined by
\begin{equation}
\hat{P}_K(s,t)=\langle\ee^{-s K}\rangle
\end{equation}
and the related cumulant generating function
\begin{equation}
\psi_K(s)=\lim_{t\to\infty}\frac{\ln \hat{P}_K(s,t)}{t}
\end{equation}
which turns out to be the largest eigenvalue of the operator $\mathbb{W}_K(s)$
defined by (see (\ref{eqn:operatorA})).
\begin{equation}
\mathbb{W}_K(s;{\cal C},{\cal C}')=\ee^{-s} W({\cal C}'\to{\cal C})-
r({\cal C})\delta_{{\cal C},{\cal C}'}
\end{equation}

There are a number of ways to obtain $\psi_K(s)$ in the regimes of interest
$s\to 0^\pm$ and $s\to \pm\infty$. The results are summarized here, while
technical details will be published elsewhere.
All these results are valid in the limit of large systems.

\noindent
$\bullet$ In the limit $s\to -\infty$
\begin{equation}
\lim_{L\to\infty} \psi_K(s)/(DL)=2\ee^{-s}\frac{\sin\pi \rho}{\pi}-2\rho(1-\rho)-
2\frac{\sin^2(\pi\rho)}{\pi^2}+{\cal O}(\ee^s)
\end{equation}
This result relies on a mapping to weakly interacting fermions, by means of a Jordan-Wigner transformation.

\noindent
$\bullet$ 
In the limit $s\to 0^-$,
\begin{equation}\label{ep3}
\lim_{L\to\infty} \psi_K(s)/(DL)=
-2\rho(1-\rho) s+\frac{2^{7/2}}{3\pi} [\rho(1-\rho)]^{3/2}|s|^{3/2}+{\cal O}(s^2)
\end{equation}
The method that was used in this case --~relying on a Bethe ansatz~--
could not be applied to the $s\to 0^+$ case.

It is not so surprising to find a non-analytic behavior for $\psi_K$,
as the symmetric exclusion process has already revealed
non-analytic behavior for the particle current distribution
function~\cite{lebowitzspohn,kim95}.

As the first derivative of $\psi_K(s)$ is still continuous in
$s=0$, one could speak of a dynamical phase transition of order
higher than one.

\noindent
$\bullet$ 
In the limit $s\to +\infty$ 
\begin{equation}
\lim_{L\to\infty} \psi_K(z)/D=-2+z^2+{\cal O}(z^3)
\end{equation}
This $z\to 0$ behavior is quite distinct from that found by Derrida and Lebowitz~\cite{derridalebowitz} studying the TASEP,
who found $\psi_K(z)=-1+z^N$ (for $N<L/2$). This is due to the strong irreversibility of the TASEP that prohibits backward jumps to take place (if $N<L/2$, $N$ jumps, instead of $2$ for the SEP, are necessary to return to a single cluster configuration). 

\section{Physical example 3: Infinite range Ising model}
\label{subsec:Ising_spin_system}
We now turn to our next example, namely a system of  $N$ Ising spins ($N\gg 1$) $\sigma_i=\pm 1$ interacting with infinite range forces and equilibrated  at the inverse temperature $\beta$.  The Hamiltonian of this infinite range Ising model reads
\begin{equation}
 \mathcal H (\boldsymbol \sigma =\{\sigma_i\}) = -\frac{1}{2N} \sum_{i,j} \sigma_i \sigma_j
= -\frac{M^2}{2N}
\end{equation}
where $M = \sum_{i} \sigma_i$ is the magnetization.
The equilibrium probability $P_\text{eq}(\boldsymbol \sigma)$ of a spin configuration
$\boldsymbol\sigma=\{\sigma_i\}$ is given by the Boltzmann-Gibbs factor
$P(\boldsymbol\sigma)\propto \exp[-\beta \mathcal H(\boldsymbol\sigma)]$.  In order to
describe its time dependent and chaotic properties we endow the system with a
continuous time Glauber-like dynamics in which each spin $\sigma_i$ flips
with a rate
\begin{equation}
  W(\sigma_i \rightarrow -\sigma_i) = \ee^{- \beta \sigma_i \frac{M}{N}}
\end{equation}
This is precisely the evolution rule considered by Ruijgrok and Tjon~\cite{ruijgroktjon} who first studied the dynamics of this system.
The rate of escape from a configuration with magnetization $M$ depends only on $M$
and is equal to
\begin{equation}
r(M)=N\cosh\frac{\beta M}{N}-M\sinh\frac{\beta M}{N}
\label{eqn:eqr}
\end{equation}

The master equation can be cast in the form of an evolution equation for the state vector $|\Psi\rangle=\sum_{\boldsymbol\sigma}P(\boldsymbol\sigma,t)|\boldsymbol\sigma\rangle$, 
\begin{equation}
\frac{\dd|\Psi\rangle}{\dd t}=\mathbb{W}|\Psi\rangle
\end{equation}
where
\begin{eqnarray}
\mathbb{W} & = & \sum_j \left[\sigma_j^x -1\right]
\ee^{- \beta \sigma_j^z \frac{M^z}{N}} \\
& = & \left(M^x-N\right)\cosh\frac{\beta M^z}{N}+\left(M^z + iM^y\right) \sinh\frac{\beta M^z}{N} \label{eqn:eqW}
\end{eqnarray}
Here, the evolution operator $\mathbb{W}$ is expressed in terms of spin $N$
 operators $M^\alpha$, $\alpha=x,y,z$ ($\sum_\alpha M^{\alpha 2}=N (N+2)$), with
$M^\alpha = \sum_j \sigma_j^\alpha$ (tensor products are
implied for the Pauli matrices $\sigma_j^\alpha$ acting on site $j$).
Note that this expression is obtained under the assumption that
the probability $P(\boldsymbol\sigma,t)$ depends only on the magnetization,
which is the case in particular for the stationary state.

An alternative way to describe the system would be to follow
another Markov variable than the full configuration $\boldsymbol\sigma$,
such as the global magnetization $M$. That $M$ is a Markov variable is of course an artifact of our mean-field model. One is now interested in the evolution equation for the magnetization state vector 
$|\Psi^{(M)}\rangle=\sum_{M}P(M,t)|M\rangle$.
It should be noted that (\ref{eqn:eqW}) gives the evolution operator for the
state vector $|\Psi^{(M)}\rangle$
(with the $M^\alpha$ taken as operators acting on magnetization states),
only if the states $|M\rangle$ are defined by
\begin{equation}
|M\rangle = \sum_{\boldsymbol\sigma}{\binom{N}{\frac{N+M}{2}}}^{-1}\delta\left[M - \sum_i \sigma_i\right]|\boldsymbol\sigma\rangle.
\end{equation}

For the somewhat more intuitive definition
\begin{equation}
|M\rangle = \sum_{\boldsymbol\sigma}\delta\left[M - \sum_i \sigma_i\right]|\boldsymbol\sigma\rangle,
\end{equation}
the evolution operator would rather be
\begin{eqnarray}
\mathbb{W} & = & \frac{M^x-iM^y}{2} \frac{N+M^z}{N-M^z+2} \ee^{-\beta \frac{M^z}{N}}
\nonumber\\
& + & \frac{M^x+iM^y}{2} \frac{N-M^z}{N+M^z+2} \ee^{+\beta \frac{M^z}{N}}
 - r(M^z)
\end{eqnarray}
as the escape rate from a given state $|M\rangle$ is still $r(M)$.
In the following, we shall always refer to the description by the
full spin-state $|\Psi\rangle$, except when stated otherwise.

We now turn to the topological pressure.
\subsection{Topological pressure -- paramagnetic state}
The topological pressure is the largest eigenvalue of the operator $\mathbb{W}_+$ whose expression reads
\begin{equation}
\mathbb{W}_+=M^x\cosh\frac{(1-s)\beta M^z}{N}r(M^z)^s+iM^y\sinh\frac{\beta(1-s) M^z}{N}r(M^z)^s-r(M^z).
\end{equation}
In the high temperature phase, it suffices to resort to the same Holstein-Primakoff representation of the spin operator as that used in \cite{ruijgroktjon},
\begin{equation}\label{defaa+}
M^x=N-2 a^\dagger a,\; M^y=-i\sqrt{N}(a^\dagger-a),\; M^z=\sqrt{N}(a^\dagger+a)
\end{equation}
in order to write $\mathbb{W}_+$ as a free boson operator whose ground-state energy yields the following topological pressure
\begin{equation}\label{psiplussigma}
{\psi}_+(s)=N(N^s-1)+N^s(1-(1-s)\beta)-N^{s/2}\sqrt{N^s(1+s\beta(2-\beta))-\beta(2-\beta)}
\end{equation}
It is also possible to compute the large deviation function associated with
the observable
\begin{equation}
Q_M = \sum_{n=0}^{K-1}\ln \frac{W(M_n\to M_{n+1})}{r(M_n)}
\end{equation}
and we find
\begin{equation}\begin{split}
\label{psiplusM}
{\psi}_M(s)  =& \lim_{t\to\infty}\frac{1}{t}\ln\langle \ee^{-s Q_M}\rangle\\
=&  (2^s-1)N+2^s(1-s)(1-\beta)\\
& -2^{s/2} \sqrt{2^s(1-s(1-\beta)^2)-\beta(2-\beta)}
\end{split}\end{equation}
We remark that in (\ref{psiplussigma}) (resp. (\ref{psiplusM})), to leading order, the distribution
of $Q_+$ (resp. $Q_M$) is a Poisson law of parameter $\ln N$ (resp. $\ln 2$), which reflects the completely randomized nature of the paramagnetic state. These results are valid in the high temperature $\beta<1$ phase. We now address the ordered state.
\subsection{Topological pressure -- ferromagnetic state}
It appears that below the critical temperature, the topological pressure 
shows much more complex features. The most notable of them is that the $Q_+$ observable ceases to be Poisson distributed even to leading order in the system size. This is at variance with what has been observed for the paramagnetic state.
In order to gain some insight into the difference between the high and
low temperature behaviors,
we decompose the fluctuating magnetization $M(t)$ into
\begin{equation}
  M(t) = N m +\xi(t) \sqrt{N}
\end{equation}
This defines the noise strength $\xi(t)$, which we expect to remain of order unity. The fluctuating escape rate from a configuration with magnetization $M$ given by
(\ref{eqn:eqr}) is, for $N$ large, given by
\begin{equation}\label{simplicite}
  r(M) = N\sqrt{1-m^2} - \xi\, \frac{m}{\sqrt{1-m^2}} \sqrt{N}
         + \frac{1}{2} \xi^2 \beta 
           \left( \beta - \frac{2}{1-m^2}\right)
        +{\cal O}(1/\sqrt{N})
\end{equation}
where the mean-field equation of state $\tanh(\beta m) = m$ was used.
From (\ref{simplicite}) we see that $r(M)$ shows only ${\cal O}(1)$ fluctuations
at $\beta<1$ ($m=0$), rather than the generically expected ${\cal O}(\sqrt {N})$
fluctuations.  Fluctuations in the high temperature regime are thus much lower than
in the ordered state. This will lead to more tedious mathematics in the ordered phase.

Before tackling these, an interesting way to pinpoint the different nature of the high and low temperature phases is to inspect first a simpler quantity, namely the number $K(t)$ of magnetization changes that have occurred over a time interval $[0,t]$.
It should be noticed that the value of $K$ is the same, whether we describe
the system by its full configuration $\boldsymbol\sigma$ or only by
its global magnetization $M$.

As explained in Sec.\,\ref{subsec:h_KS_R}, the large deviation function for $K$,
$\psi_K(s)=\lim_{t\to\infty}\frac{1}{t}\ln\langle\ee^{-s K}\rangle$ is the
largest eigenvalue of $\mathbb{W}_K$, which has matrix elements
\begin{equation}
\mathbb{W}_K(M\mp 2,M)=z\frac{N\pm M}{2N}\ee^{\mp\beta M/N}-r(M)\delta_{M,M\mp 2},\;\;z=\ee^{-s}
\end{equation}
We find, again using (\ref{defaa+}), above the critical temperature, that
\begin{equation}\label{psiKaboveTc}
 \psi_K(z) =  (z-1)N + z(1-\beta)- \sqrt{ z(z - \beta (2-\beta))}
\end{equation}
Note that at the critical point, $\psi_K(z)=N(z-1)-\sqrt{z(z-1)}$. The singularity has moved from $s=-\ln \beta(2-\beta)$ to $s=0$ ($z=\ee^{-s}$).
Below the critical temperature, expressing $\mathbb{W}_K$ in terms of $a$ and $a^\dagger$ as defined in (\ref{defaa+}) does not yield a free boson operator. We first quote our results and then sketch the route that has led to them. Retaining the leading terms in $N$, we find that $\psi_K$ has the following implicit expression:
\begin{equation}\begin{split}
\psi_K(z)    =&N \left[z\sqrt{1-m_K^2} - \cosh \beta m_K + m_K \sinh \beta m_K\right]\\&+z\frac{1-(1-m_K^2)\beta}{\sqrt{1-m_K^2}} + \sqrt{\phi_K(z)}
\end{split}\end{equation}
where
\begin{align}
 \phi_K(z) =&\frac{z^2}{1-m_K^2}
  \big[3-6(1-m_K^2)\beta + 4(1-m_K^2)^2\beta^2 \big]  \notag
 \\
  +& \beta z \sqrt{1-m_K^2}
  \big[ (2-\beta) \cosh \beta m_K + \beta m_K \sinh \beta m_K  \big]
  \big[ 1-2(1-m_K^2)\beta\big]
\end{align}
and $m_K(\beta,z)$ is the solution of
\begin{equation} \label{eqn:eqn_m_spins}
 m_K\beta \cosh \beta m_K + (1-\beta) \sinh \beta m_K =
 \frac{m_K}{\sqrt{1-m_K^2}} \ z
\end{equation}
such that $\psi_K(z)$ is the largest. When $z=1$, equation (\ref{eqn:eqn_m_spins}) is of course solved by the solution $m_K^{(0)}(\beta)$ of the mean-field equation
\begin{equation}m_K^{(0)} = \tanh (\beta m_K^{(0)})
\end{equation}
From that remark, it is possible to expand $\psi_K$ around $s=0$ in powers of
$s$ by searching solutions of (\ref{eqn:eqn_m_spins}) in the form 
$m_K=m_K^{(0)}+m_K^{(1)}\,s+\ldots$. Defining $c_0(\beta)=\cosh m_K^{(0)}(\beta) $, we find
\begin{align}\label{rh1}
 \frac{1}{N\,t} \langle K \rangle     &= \frac{1}{c_0} 
 + \frac{\beta}{N} \frac{c_0^2(2-3\beta)+\beta^2}{2 c_0 (c_0^2-\beta)^2}
 +{\cal O}(1/N^2) \\\label{rh2}
 \frac{1}{N\,t} \langle K^2 \rangle_c &= 
     \frac{1}{c_0} + c_0 \frac{c_0^2-1}{(c_0^2-\beta)^2} +{\cal O}(1/N)\\\label{rh3}
 \frac{1}{N\,t} \langle K^3 \rangle_c &= 
     \frac{1}{c_0} +3c_0 \frac{c_0^2-1}{(c_0^2-\beta)^5}  
     \big[ c_0^6 - (1+\beta)c_0^4-\beta(1-3\beta)c_0^2-\beta^3
     \big] +{\cal O}(1/N)
\end{align}
An interesting spinoff of this $s\to 0$ expansion is that it shows that in the low temperature phase ($c_0>1$), the number of steps  $K$ is not distributed according to a Poisson
distribution, even to leading order $N$ (if it were so, only the $1/c_0$ term in the right hand sides of (\ref{rh1},\ref{rh2},\ref{rh3}) would be present in the above cumulants).\\

The way the parameter $m_K(\beta,s)$ came out from the formalism
is the following. In the high temperature phase, the Holstein-Primakoff
representation (\ref{defaa+}) --~with no rotation~-- allows directly
to write $\mathbb{W}_+$ as a free boson operator.
In the low temperature phase, it is necessary to rotate 
the spin operators $M^x,M^y,M^z$ by an angle $\theta$ around
the $y$ axis, in order that $\mathbb{W}_K$ becomes a free boson
operator, with $\sin\theta=m_K(\beta,s)$ (after a suitable additional
$\theta$-dependent rotation around the $z$ axis).

It is intriguing
that by expressing the escape rate $r(M=Nm_K)=N(\cosh(\beta
m_K)-m_K\sinh\beta) m_K$ as a function of $p$ through
$m_K(p)=\sqrt{1-p^2}$, one can see, by exploiting
(\ref{eqn:eqn_m_spins}), that to leading order
$\frac{1}{N}\psi_K(z)=\text{max}_p\{z p-\frac{1}{N}r(p)\}$ (a
property holding in the $\beta<1$ phase as well).

\begin{figure}[htbp]
  \centering
  \includegraphics[width=0.7\columnwidth]{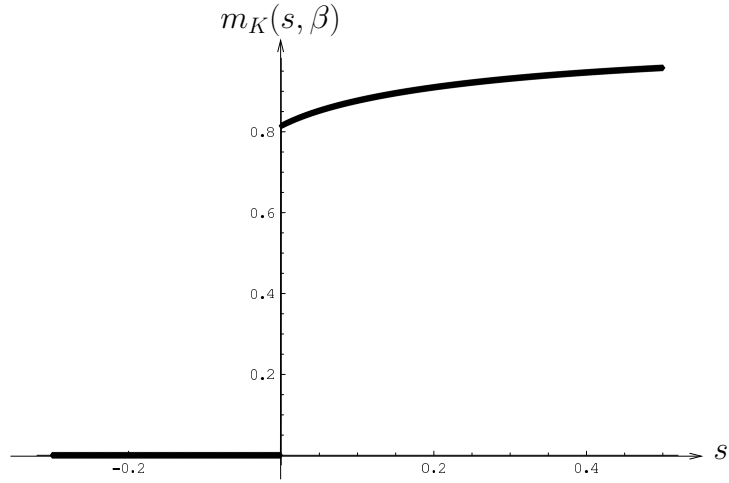}
  \caption{Plot of the rotation parameter $m_K(s,\beta)$ as a function of $s$ at $\beta=1.4$. The jump discontinuity at $s=0$, in finite size $N$, is smoothened into a continuous but steep drop centered around a critical value $s_c={\cal O}(N^{-1})<0$.}
  \label{menfctdes}
\end{figure}

\begin{figure}[htbp]
\centering
\includegraphics[width=0.7\columnwidth]{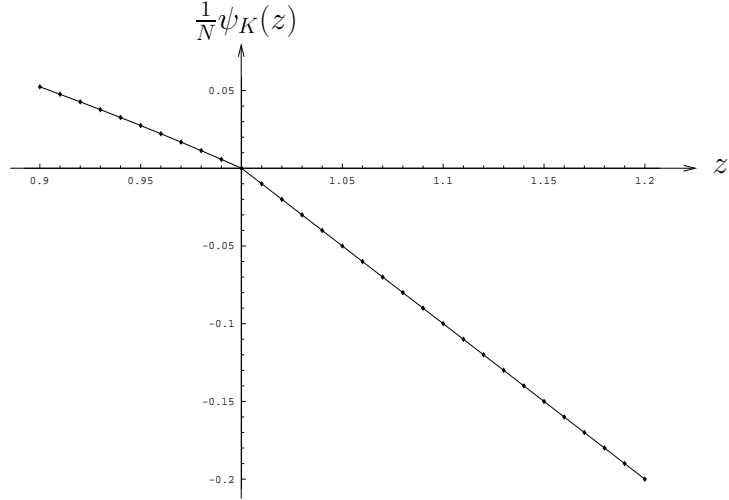}
\caption{Plot of $z\mapsto \lim_{N\to \infty} \psi_K(z)/N$ at $\beta=1.4$, with $z=\ee^{-s}$. The first derivative is discontinuous at $z=1$.}
\label{psiKbeta=1.4}
\end{figure}

The physical meaning of this $m_K(\beta,s)$ is interesting in itself: in order to
 arrive at an expression for the evolution operator involving free bosons, one
 must be describing its low lying excitations, which requires knowing its
 ground-state eigenfunction (the state $\tilde{P}_K(M,s)$ appearing in (\ref{vectpropreA})
 that has the eigenvalue $\psi_K(s)$). In the high temperature phase, the
 average magnetization restricted to histories with a prescribed value of
 $K$ is zero, because forcing a given value of $K$ does not force the system
 into the broken phase. However, in the broken phase, the nonzero magnetization
 is itself a weighted average of average magnetizations corresponding to various
 values of $K$, and there is no reason for each value of $K$ to contribute
 equally to $m_K^{(0)}(\beta)$. Instead we have that
\begin{equation} \label{eq:mKofs_spins}
m_K(s,\beta) = \lim_{N\to \infty}\frac{1}{N} \sum_{M}M\tilde{P}_K(M,s)\neq m_K^{(0)}(\beta).
\end{equation}
After all, it is reasonable that histories with $K$ far from its typical value are characterized by different magnetizations. In other words, the ground state is highly nontrivial, as opposed to the high-temperature phase. In order to further illustrate our point, we have plotted $s\mapsto m_K(s,\beta=1.4)$ in Fig.\,(\ref{menfctdes}). There it can be seen that $m_K(s,\beta)$ jumps from a nonzero value at $s>0$ to zero at $s<0$. On the one hand, at $s>0$ one is probing the regime in which $K/t$ is typically smaller than its average value $\langle r(M)\rangle $, which we expect to be more frozen than typical configurations, that is, more ordered: this accounts for $m_K(s,\beta)$ growing with $s$. On the other hand, at $s<0$, one is selecting histories that have a typical $K/t$ larger than average, so that the corresponding states should be less ordered. There is in fact a dynamical first order phase transition as $s$ varies from $0^+$ to $0^-$, where $m_K(\beta,s)$ jumps from a nonzero value to 0, which corresponds to a paramagnetic state.  
The jump discontinuity of $m_K(\beta,s)$ yields a discontinuity in the derivative of $\psi_K$ (which itself, being convex, must be continuous) as shown in Fig.\,(\ref{psiKbeta=1.4}), which reads
\begin{equation}
\left.\frac{\dd}{\dd s}\left[\lim_{N\to\infty}\frac{\psi_K(s)}{N}\right]\right|_{0^+}-\left.\frac{\dd}{\dd s}\left[\lim_{N\to\infty}\frac{\psi_K(s)}{N}\right]\right|_{0^-}=\sqrt{1-{m_K^{(0)}}^2}
\end{equation}
where $m_K^{(0)}=m_K(\beta,0)$ is the solution to $m_K^{(0)}=\tanh\beta m_K^{(0)}$. For finite $N$, both derivatives are equal to  $\sqrt{1-{m_K^{(0)}}^2}$.

\begin{figure}[htbp]
\centering
\includegraphics[width=0.7\columnwidth]{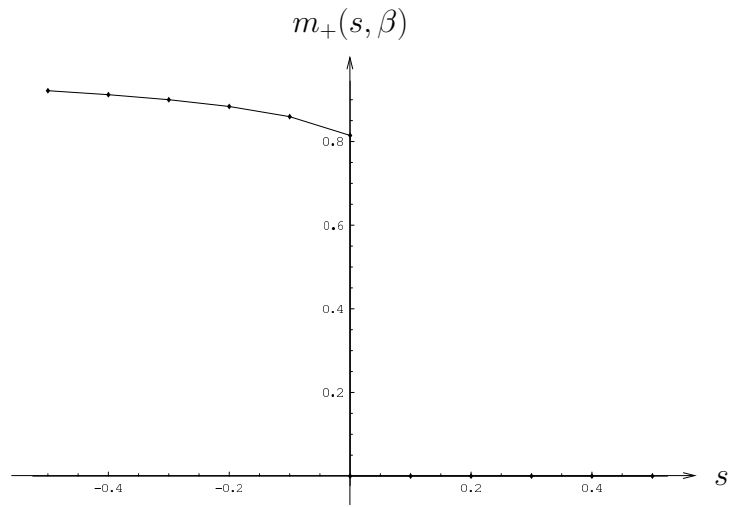}
\caption{Plot of $s\mapsto m_+(s,\beta)$ at $\beta=1.4$ in the limit of large systems.}
\label{m+beta=1.4}
\end{figure}

Returning now to the topological pressure, we parallel the reasoning carried out previously for $K$ in terms of $Q_+$: we write the corresponding operator $\mathbb{W}_+$ and perform a rotation of the magnetization operators $M^\alpha$ (the angle $\theta$ involved is such that $\sin\theta=m_+(\beta,s)$) in order that it can be expressed in terms of free bosons, and we find that below the critical temperature
\begin{equation}
 \psi_+(s)   =  N \psi^{(N)}(s)+ \psi^{(0)}(s)
\end{equation}
where the order $N^1$ and order $N^0$ coefficients are given by
\begin{align}
 \psi_+^{(N)}(s)&= p^{2-s} q^s - \frac{q}{2}\\
 \psi_+^{(0)}(s)&= q^s\, (1-s) \,\big(1-p^2\beta\big)\, p^{-(1+s)} 
 - \sqrt{\Delta_0 + \left(\frac{q}{p}\right)^s\,\Delta_1+\left(\frac{q}{p}\right)^{2s}\,\Delta_2}
\end{align}
where we used the notation
\begin{align}
  p&=\sqrt{1-m_+^2}\\
  q&=2 \big(\cosh \beta m_+ - m_+ \sinh \beta m_+\big)\\
  \overline q&=2 \big(\sinh \beta m_+ - m_+ \cosh \beta m_+\big)\\
  \Delta_0 &= -\big[ \beta m_+ \cosh \beta m_+ +(1-\beta)\sinh \beta m_+\big]^2\\
  \Delta_1 &= - p \beta \, 
               \big[(2-\beta) \cosh \beta m_+ +m_+ \beta \sinh \beta m_+ \big]  \\[1mm]
  \Delta_2 &= \frac{2}{p^2}-1 + \frac{4s}{p^2 q^2} 
              \left[\tfrac{1}{2}m_+ q\overline q -
                   p^2(1-p^2\beta)^2    \right]
\end{align}
and the rotation parameter $m_+(s,\beta)$ is the solution of
\begin{align} \label{eqn:eqn_m_spins_psiplus}
p^{1+s} \, \big[ \beta m_+ \cosh \beta m_+ +(1-\beta)\sinh \beta m_+\big]
   &= q^{s-1} \big[m_+q+s\overline q(1-p^2\beta)\big]
\end{align}
such that $\psi_+(s)$ is the largest. Again the quantity $m_+(s,\beta)$ has the meaning of an average magnetization biased, for $s\neq 0$, over histories that are more (resp. less) random than the typical history for $s>0$ (resp. $s<0$). For that reason we expect $m_+(s,\beta)$ to be a decreasing function of $s$, as is confirmed by plotting $m_+(s,\beta)$ obtained from (\ref{eqn:eqn_m_spins_psiplus}) as a function of $s$ for $\beta>1$, see Fig.\,(\ref{m+beta=1.4}).
Trajectories split into two classes, ordered and disordered ones.
This splitting is not present in the unbroken phase ($\beta < 1$),
for which $m_+(\beta<1,s)=0$ $\forall s$.

\subsection{Kolmogorov Sinai entropy and chaoticity}
\label{subsubsec:h_KS_P_spins}
\begin{figure}[htbp]
  \centering
  \includegraphics[width=0.7\columnwidth]{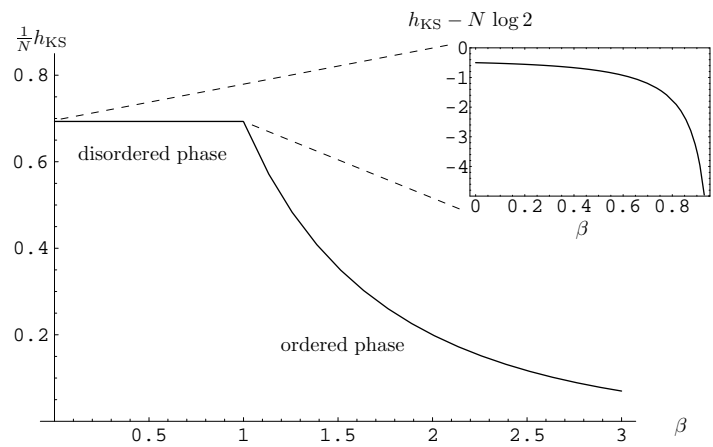}
  \caption{Kolmogorov-Sinai entropy $h_\KS$ in the stationary state, 
    as a function of $\beta$. In the ordered phase ($\beta > 1$), the variations of $h_\KS$ are
    of order $N$, while in the disordered phase, they are of order $1$ (inset). }
  \label{fig:h_KS_of_beta}
\end{figure}

Here we focus on the KS entropy related to the process $M(t)$
 --~and defined as before from the $Q_M$ observable~--,
which is luckily extensive.
In the stationary state, $h_\KS$ (in magnetization space) depends on $\beta$ through $c=\cosh \big[\beta\, m(\beta)\big]$ where $m(\beta)$ is the solution of the mean field equation
\begin{equation}
   \lim_{N\to \infty} \frac{1}{N} h_\KS = \left\{ \begin{aligned}
       \ln 2 & \quad \text{if}\quad \beta < 1 \\
      \frac{1}{c} \ln 2 & \quad \text{if} \quad \beta > 1
    \end{aligned}\right.
\end{equation}
To follow how $h_\KS$ depends on $\beta$ in the high temperature phase ($\beta <1$) one has to expand up to order  $0$. We find
\begin{equation}
h_\KS - N \ln 2= - \frac{1+(\ln 2-1 ) \beta (2-\beta)}{2(1-\beta)}
\end{equation}
Results are shown in Fig.\,(\ref{fig:h_KS_of_beta}).
In a state $P$ of average magnetization $m$, $h_\KS[P]$ only depends, to leading order in $N$, on $m$.
\begin{equation}
    \frac{1}{N} h_\KS[P] =
     \ee^{\beta m} \frac{1-m}{2} \ln 
     \left[ 1+\frac{1+m}{1-m}\ee^{-2\beta\, m} \right]
      \ + \
     \ee^{-\beta m} \frac{1+m}{2} \ln 
     \left[ 1+\frac{1-m}{1+m}\ee^{2\beta\, m} \right]
\end{equation}
In a similar way
\begin{equation}
    \frac{1}{N} h^R_\KS[P] =
     \ee^{\beta m} \frac{1-m}{2} \ln 
     \left[ 1+\frac{1-m}{1+m}\ee^{2\beta\, m} \right]
      \ + \
     \ee^{-\beta m} \frac{1+m}{2} \ln 
     \left[ 1+\frac{1+m}{1-m}\ee^{-2\beta\, m} \right]
\end{equation}

\begin{figure}[htbp]
  \centering
  \includegraphics[width=0.7\columnwidth]{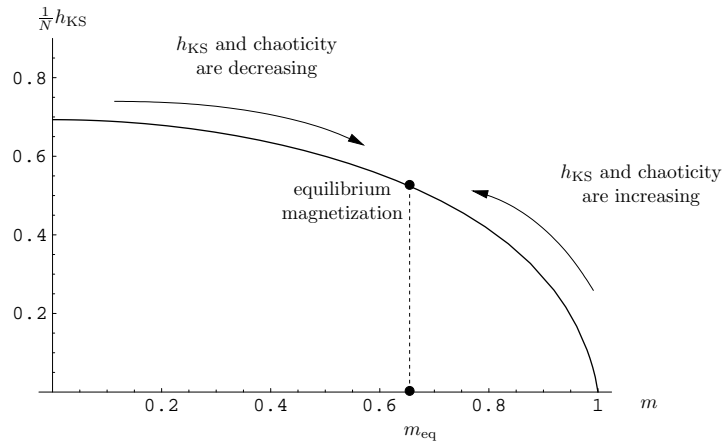}
  \caption{Kolmogorov-Sinai entropy $h_\KS[P]$ in a state $P$ of mean
    magnetization $m$ at fixed $\beta=1.2$.  }
\end{figure}

\begin{figure}[htbp]
  \centering
  \includegraphics[width=0.7\columnwidth]{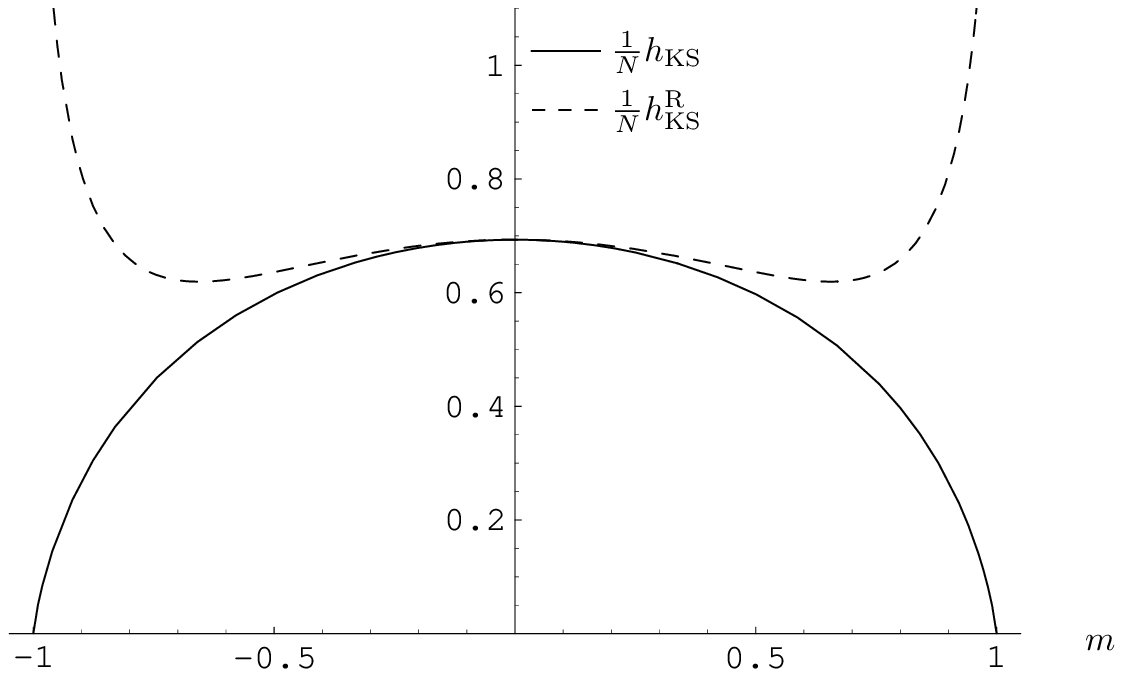}
  \caption{Direct and time-reversed Kolmogorov-Sinai entropies in a state of 
    mean magnetization~$m$, in the disordered phase ($\beta=0.8$).  Notice
    that, as expected, $h_\KS\leq h_\KS^R$. These two dynamical entropies are
    equal only at equilibrium magnetization~$m_\text{eq}=0$.  }
\end{figure}

\begin{figure}[htbp]
  \centering
  \includegraphics[width=0.7\columnwidth]{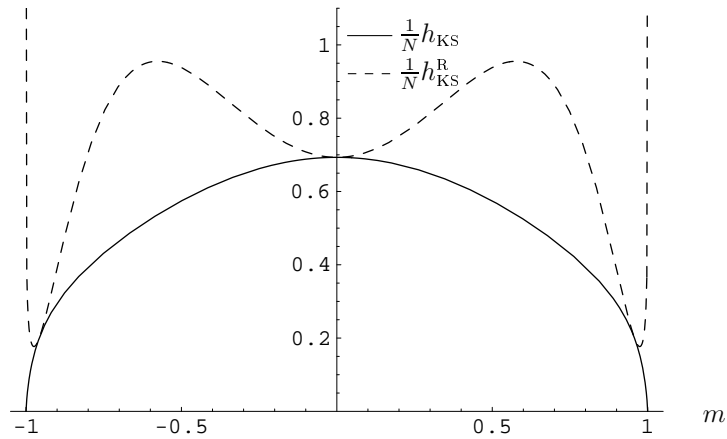}
  \caption{Direct and time-reversed Kolmogorov-Sinai entropies in a state of
    mean magnetization~$m$, in the disordered phase ($\beta=2$). Notice that,
    as expected, $h_\KS\leq h_\KS^R$. These two dynamical entropies are equal
    only at equilibrium magnetization~$m_\text{eq}\simeq \pm 0.956$ or at
    $m=0$, which is unstable.}
\end{figure}

\subsection{On the extensivity of the KS entropy}

\begin{figure}[htbp]
\centering
\includegraphics[width=0.7\columnwidth]{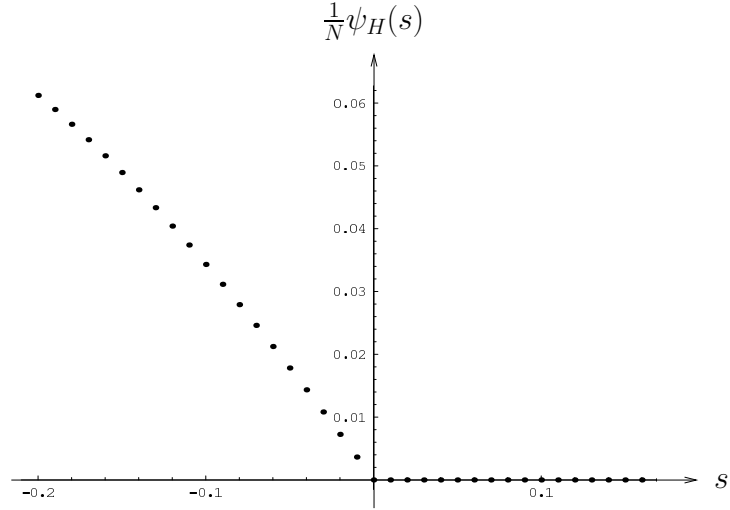}
\caption{Plot of $s\mapsto \lim_{N\to \infty} \psi_H(s)/N$ at $\beta=1.4$.}
\label{plotpsiHplus}
\end{figure}

Though the $h_\KS$ associated with the observable $Q_M$
and calculated in the previous section was luckily extensive,
in general the $h_\KS$ defined by (\ref{eqn:h_KSlong})
is not extensive in the number of degrees of freedom.
Indeed, the dominant order for $h_\KS= \psi_+'(0)$ obtained from
(\ref{psiplussigma}) reads $h_{\KS} \sim N\ln{N}$. 
By contrast, in a dynamical system, the Lyapunov spectrum, and
the KS entropy, are extensive in the number of degrees of freedom.
The nonextensivity of the $h_\KS$ calculated in this paper
was already briefly commented upon in Sec.\,\ref{Extensivity}.
As this was pointed out in Sec.\,\ref{sec:taseprand},
it is not specific to continuous time.
Still, we wish here to suggest some possible cures.
In order to obtain an extensive topological pressure, we may scale the probability of a step from $\C\to\C'$ with the number of available degrees of freedom. In the case of our Ising system, we introduce the observable
\begin{equation}
H=\sum_{n=0}^{K-1}\ln\frac{NW(\C_n\to\C_ {n+1})}{r(\C_n)}
\end{equation}
Note that the associated large deviation function $\psi_H(s)=\lim_{t\to\infty}\frac{1}{t}\ln\langle\ee^{-s H}\rangle$, in spite of remaining convex, will {\it a priori} no longer be a monotonously increasing function of $s$, which is a defining property of a R\'enyi entropy. Skipping technical details, we have found that in the high temperature phase
\begin{equation}
\psi_H(s)=1-\beta(1-s)-\sqrt{(1-s)(1-\beta)^2+s},\;\;\psi_H(s)\stackrel{\beta=1}{=}s-\sqrt{s}
\end{equation}
which has a trivial thermodynamic limit $\psi_H/N\to 0$ as $N\to\infty$. On the other hand, for $\beta>1$, we obtain
\begin{equation}
\frac{\psi_H(s)}{N}=(r(N m)/N)-p (r(Nm )/N)^s,\;p=\sqrt{1-m^2}
\end{equation}
and with $m$ solution to
\begin{equation}
m (r(Nm)/N)^s-mp\beta\cosh\beta m-p(1-\beta)\sinh\beta m=0
\end{equation}
Combining these results into a single plot, Fig.\,(\ref{plotpsiHplus}), shows that some features present in $\psi_+$ (such as $\psi_+(s>0)=0$) are unaffected in $\psi_H(s)$:
both are monotonous (only to leading order in $N$ for $\psi_H(s)$), and non-analytic in $s=0$. 
We do not have further argument in favor of using $\psi_H/N$ as a {\it bona fide} topological pressure but that it is simple and that it seems to be sharing similar properties as the original $\psi_+$, at least for the model at hand (yet it can be shown that $\psi_H'(0)\leq 0$, i.e. opposite sign to $h_\KS$).

\subsection{One dimensional Ising model}

We consider an Ising chain of $N$ spins in contact with
a thermal bath at inverse temperature $\beta$. The energy writes
\begin{equation}
 H = - \sum_i \sigma_i \sigma_{i+1}
\end{equation}
We endow the system with periodic boundary conditions and
Glauber dynamics with spin flip rate
\begin{equation}
 W_i(\sigma) = 1-\frac{1}{2}\ \gamma\ \sigma_i (\sigma_{i-1}+\sigma_{i+1})
 \qquad\text{where}\qquad
 \gamma=\tanh 2\beta
\end{equation}
The  Kolmogorov-Sinai entropy is
\begin{equation}
 h_{KS}=\left\langle 
        \sum_i W_i(\sigma) 
        \ln \frac {W_i(\sigma)}
                   {r(\sigma)}
        \right\rangle
\end{equation}
In the limit $N\rightarrow\infty$
\begin{equation}
 h_{\KS}=
 \frac{N\ln N}{\cosh 2\beta} + N
 \big[
   2\beta \gamma \tanh^2 \beta\ - (1+\gamma^2)
   \ln\cosh 2\beta
 \big]
 + 2 \sinh^2 \beta 
 + {\cal O}(\ln N/N)
\end{equation}
which is computed using that the correlations read $\langle\sigma_i\sigma_{i+r}\rangle=\tanh^{r}\beta$. It is, as expected, an increasing function of temperature.
\section{Physical example 4: Contact Process}
\label{subsec:contact_process}
\subsection{Motivations}
We now turn our attention to the infinite-range contact process: each vertex $i$ of a fully connected graph of $N$ vertices is either empty ($n_i=0$) or occupied by a
particle ($n_i=1$). The system is endowed with a Markov dynamics with rates
\begin{equation} \label{eq:transition_rates_excl_proc}
 \left\{
 \begin{aligned}
   W\big( n_i = 1 \rightarrow n_i=0\big) & = 1  
   \\
   W\big(n_i = 0 \rightarrow n_i=1 \big) &=  
                               \lambda n/N 
 \end{aligned}
 \right.
\end{equation}
where $n=\sum_i n_i$ is the total number of occupied sites. This model has recently resurfaced in the literature: Dickman and Vidigal~\cite{dickmanvidigal} studied in detail one of its defining properties, namely that it exhibits a nonequilibrium phase transition from an active to an absorbing state as the branching rate $\lambda$ is decreased below a critical value $\lambda_c=1$, with, in finite size, a single stable state, the absorbing one. Therefore, the stationary state distribution in the active state is only quasi-stationary. The lifetime of the active state, in finite size, was studied by Deroulers and Monasson~\cite{deroulersmonasson}, who also designed a systematic way to implement finite connectivity effects. Broadly speaking, the contact process phase transition belongs to the directed percolation universality class, and as such, is the paradigmatic model of nonequilibrium phase transitions. Our motivation for looking at the contact process with our own tools is precisely the existence of a phase transition, unlike any equilibrium one, that is encountered in many guises in the literature (see Hinrichsen for a review \cite{hinrichsen}). Interestingly, absorbing state transitions are now invoked within the framework of the glass transition~\cite{schonmann, reitermauchjackle, toninellibirolifisher}. At the moment we do not wish to address refined critical properties, and we shall be content with a mean-field version that will enable us to get the global picture of how phase space trajectories are affected by the presence, in the stationary state phase diagram, of an absorbing state transition.\\

Much like the global magnetization in the infinite-range Ising model, the total number of particles $n(t)=\sum_i n_i=0,\ldots,N$ is also a Markov process, with the following rates
\begin{equation}
 \left\{
 \begin{aligned}
   W(n \rightarrow n-1) &= n
   \\
   W(n \rightarrow n+1) & = (N-n)\lambda n/N 
 \end{aligned}
 \right.
\end{equation}
As a reminder, we first sketch the main properties of the stationary state. For 
finite $N$, there is a single stationary state: this is the absorbing state where all sites are empty. The time evolution of the mean number of particles reads
\begin{equation} \label{eqn:mean_n_contact_P}
  \frac{\dd \langle n \rangle}{\dd t}  =
  \left\langle (N-n)\frac{\lambda n}{N}-n\right\rangle
\end{equation}
Given the infinite range of the interactions,  the mean field hypothesis will be valid
in the thermodynamic limit ($n$ and $N$ going to infinity, $n/N\to \rho$). In
the stationary state (\ref{eqn:mean_n_contact_P}) simplifies into
\begin{equation}
  \rho \big[ \lambda (1-\rho) -1\big] = 0
\end{equation}
We conclude that there exist two regimes according to the value of $\lambda$. For $\lambda\leq 1$, the stationary state is the absorbing  state, with all sites devoid of particles, and when $\lambda>1$,
the system reaches almost certainly a quasi-stationary state with mean density 
  \begin{equation}
\label{rhostat1}
    \rho = 1-\frac{1}{\lambda},
  \end{equation}
else it is trapped in the empty state.
From here on, we shall assume $\lambda>1$ and use $\rho$ as the
only control parameter of the model. In order to circumvent the absorbing state in finite size, it is convenient to add to the original model an additional local injection process with rate $h$, 
\begin{equation}
 \left\{
 \begin{aligned}
   W\big( n_i = 1 \rightarrow n_i=0\big) & = 1  
   \\
   W\big(n_i = 0 \rightarrow n_i=1 \big) &=  
                               h + \lambda n/N 
 \end{aligned}
 \right.
\end{equation}
or
\begin{equation}
 \left\{
 \begin{aligned}
   W(n \rightarrow n-1) &= n
   \\
   W(n \rightarrow n+1) & = (N-n)\left[h+\lambda n/N\right] 
 \end{aligned}
 \right.
\end{equation}
The stationary state becomes unique for $N\to\infty$, and the steady-state density $\rho$ is given by
\begin{equation} \label{eqn:mean_field_h}
\lambda \rho + h = \rho/(1-\rho)
\end{equation}
For $h>0$, explicit results will be expressed in terms of $\rho$ and $\lambda$.
We now want to determine the large deviation functions of $K(t)$, the number of configuration changes that have occurred over a time interval $[0,t]$, and of $Q_+(t)$, which gives access to the topological pressure.
\subsection{Special point $\lambda = \frac{2\rho-1}{(1-\rho)^2}$}\label{specialpoint}
Our first paragraph deals with a special point in parameter space that, to the best of our knowledge, has never been commented upon in the existing literature, but whose mathematical structure is extremely simple.  We decompose the total number of particles into an average and a fluctuating part,
\begin{equation}
  n = N\rho+\xi \sqrt{N} 
\end{equation}
and we express the fluctuating rate of escape from a  configuration with $n$ particles, in the stationary state.
In the absence of particle injection ($h=0$), and replacing
 $\lambda$ by its expression (\ref{rhostat1})
in terms of the stationary state density $\rho$,  we arrive at
\begin{equation}
  r(n) = 2 \rho N + \xi \frac{2-3\rho}{1-\rho} \sqrt{N}
       - \xi^2 \, \frac{1}{1-\rho}
\end{equation}
hence the special point $\rho=2/3$ (or equivalently $\lambda=3$) at which this escape rate has relative fluctuations of order ${\cal O}(N^{-1})$ that are much weaker than  the generically expected ${\cal O}(N^{-1/2})$. A similar phenomenon occurs for  $h>0$, using (\ref{eqn:mean_field_h}),
\begin{equation}
  r(n) = 2 \rho N + \xi (1-\rho) (\lambda - \Lambda) \sqrt{N}
       - \lambda \xi^2 
\end{equation}
where
\begin{equation}
  \Lambda = \frac{2\rho-1}{(1-\rho)^2}
\end{equation}
There the special point with low fluctuations
is at $\lambda=\Lambda$. Under this
constraint $\lambda=\Lambda$, the interval covered by the
stationary state density when the value of $h$ is varied
is $\frac{1}{2}<\rho<\frac{2}{3}$.
The
$\lambda=\Lambda$ behavior of $r(n)$ bears much resemblance with that
already noted for the high-temperature phase of Ising model in
(\ref{simplicite}), with the formal correspondence
$\beta<1\leftrightarrow \lambda=\Lambda$ and $\beta>1\leftrightarrow
\lambda\neq \Lambda$.
As will now be seen, huge calculational simplifications occur at $\lambda=\Lambda$.\\

The generating function for the cumulants of $K(t)$, the number of configuration changes that have occurred over a time interval $[0,t]$ is the largest eigenvalue of the 
following  operator
\begin{equation}\label{WKCP}
 \mathbb{W}_K(z) =-  \hat n+(N-\hat n)\left[\frac{\lambda \hat n}{N}+h\right] + \frac{1}{2} z 
  \left[ (M^x + i M^y)\left[\frac{\lambda \hat n}{N}+h\right] + (M^x - i M^y)  \right]
\end{equation}
where $\hat n = (N+M^z)/2$ is the particle number operator and $z=\ee^{-s}$.  Given that the detailed properties are being studied for the first time here, we shall provide the reader with a few more technical details than in the previous section on the Ising model.\\

The spectrum of $\mathbb{W}_K$ can be found
perturbatively in $N$ using the Holstein-Primakoff representation of the
magnetization operators $M^\alpha$. In general this consists in rewriting the $M^\alpha$'s as a
carefully chosen rotation of another set $L^\alpha$ of spin $N$ operators for which we will use the following exact representation in terms
of creation and annihilation operators
 \begin{align} \label{eqn:HolstPrim_x}
    L^x &= N - 2 a^\dag a\\ 
  i L^y &= a^\dag \left(N-a^\dag a \right)^\frac{1}{2} - 
                    \left(N-a^\dag a \right)^\frac{1}{2} a \\ 
    L^z &= a^\dag \left(N-a^\dag a \right)^\frac{1}{2} + 
                    \left(N-a^\dag a \right)^\frac{1}{2} a 
    \label{eqn:HolstPrim_z}
 \end{align}
The aforementioned rotation has to be chosen such that in the ground state,
$a^\dagger a$ remains small, so that an expansion can be performed.
In the present case ($\lambda=\Lambda$), we shall assume that it is already the case without any rotation,
and we shall use directly $M^\alpha=L^\alpha$.
We expand $\mathbb{W}_K$ in powers of $N$ anticipating that in the ground state $a^\dagger a$ will remain of ${\cal O}(1)$ as $N\to\infty$. And because, up to a constant contribution, $\mathbb{W}_K$ is  quadratic in terms of $a$ and $a^\dagger$ (with $N$-independent coefficients),  this is indeed the case and we find that the largest eigenvalue $\psi_K(s)$ of $\mathbb{W}_K$ has the following expression
\begin{equation}
\psi_K(z) = 
 \left\{\begin{aligned}
   \text{Root of a third degree polynomial} & \quad\text{if}\quad z<z_c \\
   2\rho\,(z-1)\, N  
   -z + \frac{z^{1/2}}{1-\rho}
   \sqrt{\rho(1-2\rho)+z(1-3\rho(1-\rho)) }
   & \quad\text{if}\quad z>z_c
  \end{aligned} \right.
\end{equation}
provided the parameters verify $\lambda =
\frac{2\rho-1}{(1-\rho)^2}$. Note that for $h=0$, that is at
$\lambda=3$, $z_c=1$ and $\psi_K(z\leq 1)=0$, while $\psi_K(z)
=\frac{4}{3}(z-1) N -z + \sqrt{z(3z-2) }$ if $ z>1$. Interestingly, to
leading order in $N$, the distribution of $K$ is a Poissonian, as was
precisely the case for the Ising model in the high-temperature
phase~(\ref{psiKaboveTc}). For $h\neq 0$, we find that for $z\to 0$
(that is for $s\to\infty$)
\begin{equation}
 \psi_K(z) = \rho \frac{2-3\rho}{(1-\rho)^2} + 
             z^2 \frac{2-3\rho}{2(1-2\rho)} +  {\cal O}(z^4)
\end{equation}
which describes reduced-activity histories with values of $K$ much smaller than $\langle K\rangle$.\\ 

\begin{figure}[htbp]
  \centering
  \includegraphics[width=0.7\columnwidth]{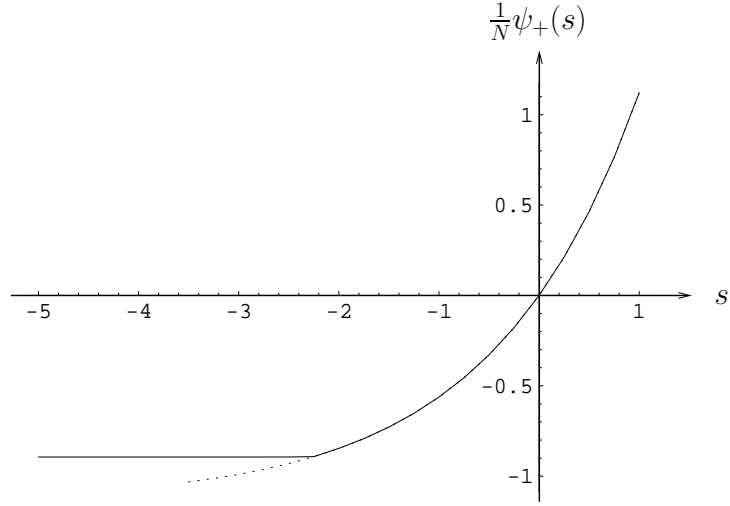}
  \caption{Topological pressure $\lim_{N\to\infty}\frac 1N
    \psi_{+}(s)$ on the special line (at $\rho=0.57$). The dashed line is
    the continuation of the strictly convex branch for $s<s_c$. }
  \label{fig:dyn_phase_tr_special_line}
\end{figure}

In much a similar way, restricting our analysis to the Markov process $n(t)=\sum_i n_i(t)$, the topological pressure $\psi_+(s)$ is the largest eigenvalue of the following operator
\begin{equation}\label{W+CP}\begin{split}
-\mathbb{W}_+(s)=& \hat n+(N-\hat n)\frac{\lambda \hat n}{N} \\
& -\frac 12 (M^x+iM^y)\left(h+\frac{\lambda\hat{n}}{N}\right)^{1-s}\left(\hat{n}+h+\frac{\lambda\hat{n}}{N}\right)^s\\
& -\frac 12 (M^x-iM^y)\left(\hat{n}+h+\frac{\lambda\hat{n}}{N}\right)^s
\end{split}\end{equation}
And again expanding the $M^\alpha$'s in powers of $N$ keeping $a$ and
$a^\dagger$ of order 1, leads to $\mathbb{W}_+$ being quadratic in $a$
and $a^\dagger$, with $N$-independent coefficients. Using the
Bogoliubov-like transformation described in Appendix~\ref{appendixA},
it is thus a simple matter to find the largest eigenvalue of
$\mathbb{W}_+$, which reads
\begin{equation}\label{psi+specialCP}
\psi_+(s) = 
  \left\{\begin{aligned}
         & 2\rho\,(2^s-1) N \ + \
           2^s(1-s) \\
         & \;\; -
           \sqrt{
                 4^s\left[\frac{1-3\rho(1-\rho)}{(1-\rho)^2}-s\right]+
                 2^s\:\frac{\rho(1-2\rho)}{(1-\rho)^2}
                } + {\cal O}(1/N)
         &\text{ if }s\geq s_c \\
         & - h N + {\cal O}(1/N)&\text{ if }s\leq s_c
         \end{aligned}
  \right.
\end{equation}
where $h = \frac{(2-3\rho)\rho}{(1-\rho)^2}$. The critical value $s_c$ that emerges in (\ref{psi+specialCP}) is given by
\begin{equation}
   s_c = \log_2 \frac{\lambda\rho}{2} 
     \ +\ 
     \frac{1}{N} \frac{1}{2 \rho \ln 2}
     \left(
            -2+\log_2 \lambda\rho + \sqrt{\lambda \rho -\log_2 \lambda \rho }
     \right)
     \ + \ 
     {\cal O}(1/N^2)
\end{equation}
When $s>0$, the expansion of the
$\mathbb{W}_+$ is valid only when $s \ll \sqrt{N}$. When $N$
is (large and) fixed, the asymptotics of $\psi_+(s)$ is
\begin{equation}
  \psi_+(s) \sim \sqrt{h N} \left( \frac{h N^2 + 2 \lambda \rho N - \lambda}{N} 
  \right)^s
 \qquad \text{ as } \quad s \to \infty
\end{equation}
In Fig.\,(\ref{fig:dyn_phase_tr_special_line}) we have plotted $\psi_+(s)$ as a function of $s$.
The most remarkable feature is the presence of dynamical transition at the critical parameter $s=s_c$. The nontrivial convex branch ceases to correspond to the largest eigenvalue of $\mathbb{W}_+$ at $s<s_c$, and it is simply replaced by a plateau. This picture, which is customary in equilibrium phase transitions, reflects the existence of an underlying first order transition. As $s$ is decreased from $s=0$ (corresponding to typical histories) one is selecting histories with less and less dynamical disorder. This indicates a phase-separation like mechanism occurring in the space of histories. We now attack the generic case for which the values of $h$ and $\lambda$ are unrestricted.
\subsection{Generating function of the number of events for any $\lambda$}
The task remains that of finding the largest eigenvalue $\psi_K(s)$ of $\mathbb{W}_K$ as given in (\ref{WKCP}). When directly expanded in $N$, the choice $M^\alpha=L^\alpha$ leads to the following expression for the evolution operator $ \mathbb{W}_K$ 
\begin{equation} \label{eqn:H_dev_N}
   -\mathbb{W}_K(z) =  
           H^{(2)} \,N 
           + \left(H_a^{(1)} a + H_{a^\dag}^{(1)}a^\dag \right)\sqrt{N}
           + \hat H^{(0)} + {\cal O}(\sqrt{N})
\end{equation}
where $H^{(2)},H_a^{(1)},H_{a^\dag}^{(1)}$ are c-numbers and $\hat H^{(0)}$ is
quadratic in $a$ and $a^\dag$. While this seems a perfectly legitimate large $N$ expansion, the presence of nonzero $H_a^{(1)}$ or $H_{a^\dag}^{(1)}$ terms in  (\ref{eqn:H_dev_N}) signals that the ground state of  $-\mathbb{W}_K$ does not correspond to the zero boson state, but rather to an ${\cal O}(N)$ boson state (on the special parameter subspace $\Lambda=\lambda$ these coefficients of the linear terms in $a$ and $a^\dagger$ somehow miraculously vanish). Indeed, in order to
compute that spectrum we need to translate the creation and annihilation
operators\footnote{for instance, through similarity transformations such as
  $\ee^{C a} a^\dag \ee^{-C a} = a^\dag +C$.} by a constant of magnitude
$\sqrt{N}$, but this mixes the whole expansion (\ref{eqn:H_dev_N}) of $\mathbb{W}_K(z)$
in powers of~$N$.  In particular, unless $H_a^{(1)} = H_{a^\dag}^{(1)}= 0$,
the truncated expansion~(\ref{eqn:H_dev_N}) is not sufficient to find the
eigenvalues of $\mathbb{W}_K(z)$, even to lowest order in $N$.  Given that we wish to describe $-\mathbb{W}_K$'s low lying excitations, with $a^\dagger a\sim{\cal O}(1)$, we must now find a way to expand around the ground-state. By contrast to Ruijgrok and Tjon~\cite{ruijgroktjon}, we
must now perform two successive rotations parametrized by $\alpha$ and $m$
(around the $y$ and the $z$ axes) of the
initial Holstein-Primakoff representation (\ref{eqn:HolstPrim_x}-\ref{eqn:HolstPrim_z}). The evolution
operator $\mathbb{W}_K(z)$ then reads
\begin{equation} \label{eqn:H_K_rotated}
 -\mathbb{W}_K(z) =  \hat n+(N-\hat n)\frac{\lambda \hat n}{N}  - \frac{1}{2} z 
  \left[ \alpha\:    (M^x + i M^y)\frac{\lambda \hat n}{N} + 
         \alpha^{-1} (M^x - i M^y)  \right]
\end{equation}
with $M^y=L^y$,
\begin{equation} \label{eqn:RuijTj_rotation}
 \begin{pmatrix} M^x   \\ M^z   \end{pmatrix} = 
 \begin{pmatrix}    p  &  -m    \\
                    m  &   p    \end{pmatrix} 
 \begin{pmatrix} L^x \\ L^z \end{pmatrix} 
 \;, \quad
  p=\sqrt{1-m^2}
 \quad \text{ and }
 \quad
 \left\{\begin{aligned}
   -1\leq m \leq 1 \\
   \alpha \geq 0 
 \end{aligned} \right.
\end{equation}
The parameters of the two rotations, $\alpha$ and $m$, will now be chosen so
that $H_a^{(1)} = H_{a^\dag}^{(1)}= 0$ in the truncated
expansion (\ref{eqn:H_dev_N}) of (\ref{eqn:H_K_rotated}).  When these
equations in $\alpha$ and $m$ have more than one solution, we have to choose
the solution which gives the highest value of $\psi_K$.  Expanding $\mathbb{W}_K(z)$ in powers of $N$ and imposing $H_a^{(1)} = H_{a^\dag}^{(1)}$ implies
that $\alpha=\sqrt{2\frac{1-\rho}{1+m}}$ and yields an expression of the
form (\ref{eqn:H_dev_N}) with
\begin{align}
 H^{(2)} &= 
   \frac{1}{4(1-\rho)} 
   \left( 
      4pz
      \sqrt{\tfrac{1}{2}(1+m)(1-\rho)}-
      (3-2\rho-m)(1+m)
   \right)\\
 H_a^{(1)} &= H_{a^\dag}^{(1)} = 
   \frac{1}{2(1-\rho)} 
   \left( 
      z(3m-1)\sqrt{\tfrac{1}{2}(1+m)(1-\rho)}+
      p(1-\rho-m)
   \right) \label{eqn:coeff_a_adag} 
\end{align}
From (\ref{eqn:coeff_a_adag}) we see that solving 
$H_a^{(1)} = H_{a^\dag}^{(1)}=0$ in $m$ leads to either $m=-1$ or 
$m$ is one of the roots of third degree polynomial. If $m=-1$ is not the
correct solution, this root must 
be inserted back into the expressions of  $H^{(2)}$ and $\hat H^{(0)}$
to get $\psi_K(s)$. With a view to avoiding further technicalities, it is  more convenient to
use algebraic elimination methods so as to find an equation on $H^{(2)}$
itself, and on the coefficients of $\hat H^{(0)}$.  Skipping details, one finds that when $m\neq-1$,
$H^{(2)}$ is one of the roots of the following polynomial
\begin{align} \label{eqn:poly_psi2}
  P(X)   & = c_3 X^3 + c_2 X^2 + c_1 X + c_0   \\
      c_3   & = 16 (1-\rho)^2  \nonumber \\
      c_2   & =
      -27\,z^4\,{\left( -1 + \rho \right) }^3 + 12\,z^2\,{\left( -1 + \rho
        \right) }^2\,\left( -4 + 3\,\rho \right) - 8\,\left( -6 + 12\,\rho -
        7\,{\rho }^2 + {\rho }^3 \right)
             \nonumber \\
      c_1 & =  
      -12\,z^4\,{\left( -1 + \rho \right) }^2\,\left( -4 + 3\,\rho \right) 
      - z^2\,\left( 96 - 228\,\rho + 184\,{\rho }^2 - 53\,{\rho }^3 + {\rho
        }^4 \right)
            \nonumber \\ & \qquad
      + {\left( -2 + \rho \right) }^2\,\left( 12 - 12\,\rho + {\rho }^2 \right)
               \nonumber \\
      c_0 & =  (1-z^2)\left[4z^2(1-\rho) -(2-\rho)^2\right]^2   \nonumber 
\end{align}
We first consider the case $s\geq 0$. In that range of $s$, we see
that by definition we must have $\psi_K(s)\leq 0$. However, the
solution $m=-1$ of $H_a^{(1)} = H_{a^\dag}^{(1)}= 0$ yields
$\psi_K(s)=0$, which is the highest possible value of $\psi_K(s)$. We
thus have $\psi_K(s)=0$ in the whole $s\geq 0$ range. We now assume
$s<0$. And again by definition we must have $\psi_K(s)\geq 0$. The
solution $m=-1$ still yields $\psi_K(s)=0$. We thus have to check
whether $P(X)$ has any negative solution. The discriminant of $P(X)$
has the simple form
\begin{align}
 \Delta &= -\frac{1}{2^{28}\: 3^3\: (1-\rho)^5} \;
          z^2\,{\left( 2 + z^2\,\left( -1 + \rho  \right)  - \rho  \right) }^2
\nonumber \\&\qquad\qquad \left\{
   3\big[24 z^2 (1-\rho)+ \rho^2+12\rho-12\big]^2  +
  {\left( 6 - \rho  \right) }^3\,\left( -2 + 3\,\rho  \right)
 \right\}^3
\end{align}
As $\Delta<0$ in the range $s<0$, $P(X)$ has three real valued roots.  Moreover, from
the coefficients of (\ref{eqn:poly_psi2}) it is easy to see that the roots of
$P(X)$ have a positive sum and a negative product, which shows that $P(X)$ has
only one negative root, namely, $H^{(2)}$. From Cardano's formula, setting
$q=(9 c_1 c_2 c_3 - 27 c_0 c_3^2 - 2 c_2^3)/(54 c_3^3)$ we
find\footnote{The expression of $\psi_K$ takes real
    values but can't be written with algebraic operations involving only real
    quantities: this is the {\em casus irreductibilis} of Cardano's formula.}
that in the range $s<0$
\begin{align} \label{eqn:psiKContactProcess}
 \psi_K^{(2)}(s) &=
  -\frac{c_2}{3\,c_3} +
  \ee^{2 i \pi / 3}
  \left(q+i \sqrt{-\Delta}\right)^{\frac{1}{3}} +
  \ee^{-2 i \pi / 3}  
  \left(q-i \sqrt{-\Delta}\right)^{\frac{1}{3}} 
\end{align}
As a remark, we notice that the two rotations of parameters $\alpha$ and $m$
 could also be understood as the
result of suitable similarity transformations of the kind $\ee^{\theta M^z}
(\ldots) \ee^{-\theta M^z}$ performed on $\mathbb{W}_K(z)$ before expanding in $N$. In
other words, finding the roots of (\ref{eqn:poly_psi2}) enabled us to perform an appropriate resummation of (\ref{eqn:H_dev_N})  to all orders in order to obtain a series whose truncation to lowest order has well defined spectrum, which makes the large $N$ expansion consistent.\\

\begin{figure}[htbp]
  \centering
  \includegraphics[width=0.7\columnwidth]{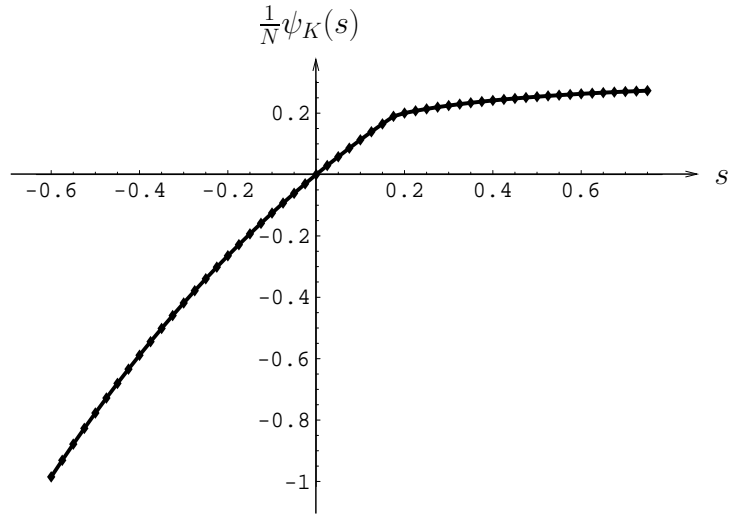}
  \caption{Plot of $\lim_{N\to\infty}\psi_K(s)$  as a function of $s$ at $\lambda=2$ and $h=0.3$. Note the presence of a jump in the first derivative at $s=s_c \simeq 0.16$.}
  \label{psiKCPlambda2h0.3}
\end{figure}

\begin{figure}[htbp]
  \centering
  \includegraphics[width=0.7\columnwidth]{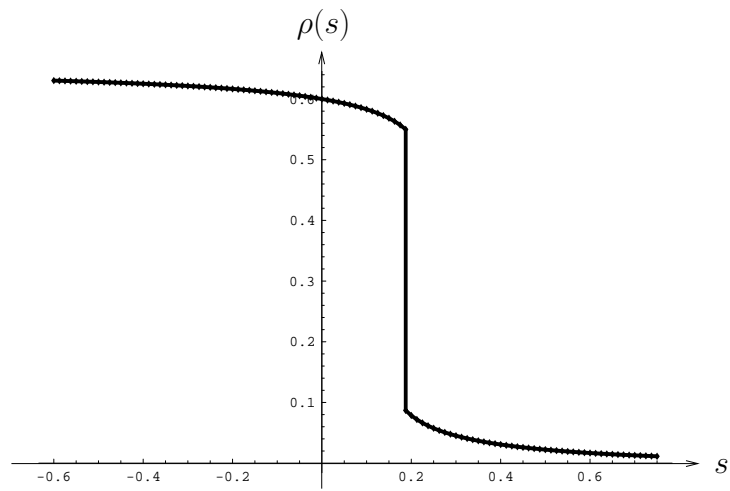}
  \caption{Plot of $\rho(s)=\frac{1+m(s)}{2}$ in the $N\to\infty$ limit as a function of $s$ at $\lambda=2$ and $h=0.3$. Note the presence of a jump at $s=s_c \simeq 0.16$.}
  \label{rhoKCPlambda2h0.3}
\end{figure}

In order to be more explicit, we now provide the limiting behavior of $\psi_K(s)=N \psi_K^{(2)}(s) + \psi_K^{(0)}(s) +{\cal O}(N^{-1})$ in two limits of interest, namely for $s\to 0^-$,
\begin{align}
 \psi_K^{(2)}(s) &= 2\rho (\ee^{-s}-1) + (2\rho-3)^2 \Big[ \frac{s^2}{\rho } 
                  - \frac{s^3\,\left( -4 + \rho \,\left( 4 + \rho  \right)  
                  \right) }{{\rho }^3} \nonumber \\
                 &\qquad 
               \frac{s^4\,\left( 432 + \rho \,\left( -1248 + 
          \rho \,\left( 1188 + \rho \,\left( -444 + 79\,\rho  \right)  \right)
                 \right)  \right) }{12\,{\rho }^5} \Big] 
          +{\cal O}(s^5)
\nonumber  
\end{align}
and for $s\to -\infty$,
\begin{align}
 \lim_{s\to -\infty} \ee^{s } \psi_K^{(2)}(s) & = 1 
\end{align}
The remaining ${\cal O}(1)$ piece in $\psi_K$ is given by
\begin{equation}
 \psi_K^{(0)}(s) = C + \sqrt{D}
\end{equation}
where $C$ is a root from the polynomial
\begin{align} P &=
z^2\,{\left( -2 + \rho  \right) }^2 + 4\,z^4\,\left( -1 + \rho  \right)  -
  2\,X^3\,{\left( -1 + \rho  \right) }^2  \nonumber \\ & -  
  X\,z^2\,\left( -1 + \rho  \right) \,\rho  + 4\,X^2\,\left( -1 + \rho  \right) \,
  \left( 3\,z^2\,\left( -1 + \rho  \right)  + 2\,\rho  \right) 
\end{align}
and $D$ is a root from the polynomial
\begin{align} P &=
256\,X^3\,{\left( -1 + \rho  \right) }^5 + 16\,X^2\,{\left( -1 + \rho  \right) }^3\,
   \left( -27\,z^4\,{\left( -1 + \rho  \right) }^2 + 24\,z^2\,\left( -1 + \rho
   \right) \,\rho  - 4\,{\rho }^2 
     \right)  \nonumber \\ &
 - z^2\,\left( {\left( -2 + \rho  \right) }^2 + 4\,z^2\,\left( -1
   + \rho  \right)  \right) \, 
   \left( 108\,z^4\,{\left( -1 + \rho  \right) }^2 + 8\,{\rho }^3 -  
     9\,z^2\,\left( -1 + \rho  \right) \,\left( -12 + \rho \,\left( 12 + \rho
   \right)  \right)  \right)  +  \nonumber \\ &
  8\,X\,z^2\,{\left( -1 + \rho  \right) }^2\,\left( -108\,z^4\,{\left( -1 +
   \rho  \right) }^2 - 8\,{\rho }^3 +  
     9\,z^2\,\left( -1 + \rho  \right) \,\left( -12 + \rho \,\left( 12 + \rho
   \right)  \right)  \right) 
\end{align}
It is now time to summarize our findings, which we do in the following
two plots Figs.\,(\ref{psiKCPlambda2h0.3}) and (\ref{rhoKCPlambda2h0.3}),
showing respectively the full plot of $\psi_K(s)$ as a function of $s$
and that of the density $\rho(s)=(1+m(s))/2$ corresponding to the
rotation parameter $m(s)$ as a function of $s$.

On Fig.\,(\ref{psiKCPlambda2h0.3}) we notice that $\psi_K(s)$ is not
analytic at some critical point $s_c$ which corresponds to the
phase-separation like mechanism depicted by the topological pressure
$\psi_+(s)$ (see Fig.\,\ref{fig:dyn_phase_tr_special_line}), but now
at the level of the number of events $K$. This result illustrates that
this simple quantity --~at least for infinite-range systems~-- already
contains much of the information given by $Q_+$ on the complexity of
histories. 

This is fully confirmed by Fig.\,(\ref{rhoKCPlambda2h0.3}).
In analogy to the Ising case~\eqref{eq:mKofs_spins}, $\rho(s)=\frac{1+m(s)}{2}$
represents the mean density in the biased state $\tilde{P}_K(n,s)$:
\begin{equation} \label{eq:rhoKofs_excl_proc}
\rho(s) = \frac{1}{N} \sum_{n}n\tilde{P}_K(n,s)
\end{equation}
As usual, at $s=0$ we recover the density in the steady state. At $s<0$ we probe
the regime in which the mean ``activity'' $K/t$ of histories is
typically larger than in the steady state.  They correspond to explored configurations where
the density is larger than the steady state density $\rho$. On the other hand, at $s>0$ histories
with smaller  $K/t$ are favored. Increasing $s$ leads to a sudden jump
in the typical density, which corresponds to a dramatic change in the kind of configurations
explored by histories with reduced activity $K/t$.

\subsection{Topological pressure: $h=0$}
\begin{figure}[htbp]
  \centering
  \includegraphics[width=0.7\columnwidth]{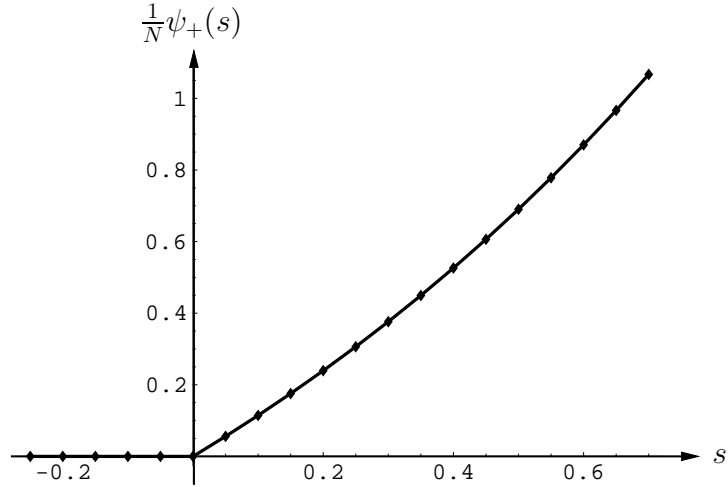}
  \caption{Plot of the topological pressure $\lim_{N\to\infty} \frac 1 N \psi_+(s)$ as a function of
    $s$ at $\lambda=5$ and $h=0$.  Note the presence of a jump in the
    first derivative at $s_c=0$.}
  \label{psiKCPlambda5h0}
\end{figure}

\begin{figure}[htbp]
  \centering
  \includegraphics[width=0.7\columnwidth]{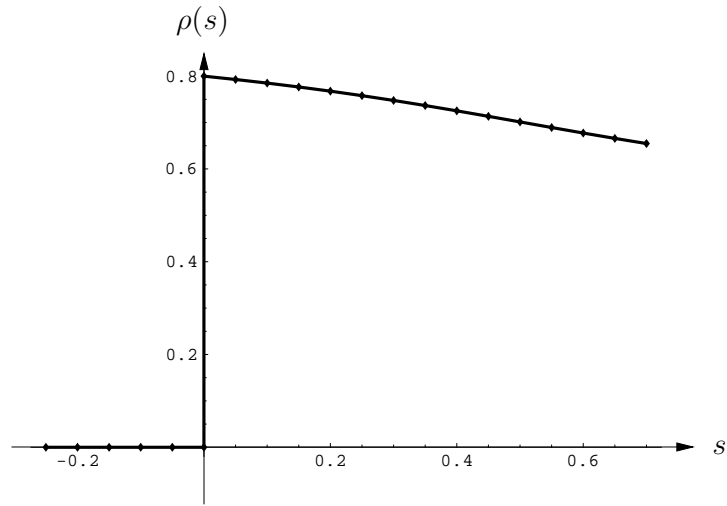}
  \caption{Plot of $\rho(s)=\frac{1+m(s)}{2}$ as a function of $s$ at
    $\lambda=5$ and $h=0$ in the large system limit $N\to\infty$. Note
    the presence of a jump at $s_c=0$.}
  \label{rhoKCPlambda5h0}
\end{figure}

We begin by attacking the $h=0$ case for which the phase diagram possesses two stationary states, the active and the absorbing one. The topological pressure $\psi_+(s)$ is the largest eigenvalue of the operator $\mathbb{W}_+(s)$ written out in (\ref{W+CP}). By techniques similar to those mentioned above, we arrive at
\begin{align}
 \psi_+(s)      &=  N \psi_+^{(N)}(s)+ \psi_+^{(0)}(s)\\
 \psi_+^{(N)}(s)&= \frac{1+m}{4}
        \left(-\frac{r}{1-\rho}+\sqrt{\frac{2q^s}{(1+m)(1-\rho)}}\right)\\
 \psi_+^{(0)}(s)&=
       (1-s)\frac{1+m}{4p}\sqrt{\frac{2q^s(1+m)}{1-\rho}}\\
& -\sqrt{
  -\frac{p}{4(1-\rho)}
           \sqrt{\frac{2q^s(1+m)}{1-\rho}}
           + \frac{q^s}{4}\frac{1+m}{1-\rho}
             \big(\Delta_0 + s\Delta_1+s^2\Delta_2\big)
}
\end{align}
where we used the notations
\begin{align}
  p&=\sqrt{1-m^2}\\
  r&=3-m-2\rho\\
  q&=\frac{r^2}{2(1-m)(1-\rho)}\\[1mm]
  \Delta_0 &=\frac{3}{2}+\frac{1}{1-m}\\[1mm]
  \Delta_1 &=\frac{2s}{r^2}\big(p^2-2r(1+m-\rho)\big)\\[1mm]
  \Delta_2 &=-\frac{(1+m)(1+m-2\rho)^2}{2(1-m)r^2}
\end{align}
The rotation parameter $m$ is the solution of
\begin{align} \label{eqn:eqn_m_spins_psiplus1}
 2 p r (1-m-\rho) =
 \sqrt{2(1-\rho)(1+m)q^s}
 \left(s(1+m)(1+m-2\rho)+(1-3m)r\right) 
\end{align}
such that $\psi^{(N)}(s)$ has the largest value. The first cumulants can be determined without toil,
\begin{align}
 \frac{1}{N\,t} \langle Q_+ \rangle     &= 
   2\rho\ln 2 -\frac{1}{4 N} 
   \left[
          \frac{\rho}{1-\rho}
          -8\frac{1-\rho}
                {\rho}
           \ln 2
   \right]
   +{\cal O}(1/N^2) \\
 \frac{1}{N\,t} \langle Q_+^2 \rangle_c &= 
   \rho (\ln 2)^2 +\frac{(2-3\rho)^2}{\rho}(\ln 2)^2
   +{\cal O}(1/N)\\
\end{align}
Fig.\,(\ref{psiKCPlambda5h0}) shows the topological pressure $\psi(s)$
and the corresponding density $\rho(s)$ is represented on Fig.\,(\ref{rhoKCPlambda5h0}).

\subsection{Topological pressure (ii): $h>0$}
Finally we turn to $h>0$ for which the explicit formulas read
\begin{align}
 \psi(s)      &=  N \psi^{(N)}(s)+ \psi^{(0)}(s)\\
 \psi^{(N)}(s)&= \frac{1}{4}
        \left(- \overline r + 2 p \sqrt{r\,q^s}\right)\\
 \psi^{(0)}(s)&=\frac{(1-s) u}{2p}\sqrt{\frac{q^s}{r}}
  - \sqrt{q^s\,\big(\Delta_0+s\Delta_1+s^2\Delta_2\big) }
\end{align}
where we used the notations
\begin{align}
  p&=\sqrt{1-m^2}\\
  r&=2\big(2h+\lambda(1+m)\big)\\
  \overline r&=\tfrac{1}{2}(1-m)r+2(1+m)\\
  q&=\frac{\overline r}{r p^2}\\[1mm]
  u&=4h+\lambda(1+m)^2\\[1mm]
  \Delta_0 &=\frac{\lambda^2p^2}{4r}+\frac{h}{p^2}
             +\frac{\lambda}{2}\frac{p^2+m}{1-m}\\[1mm]
  \Delta_1 &=
    4\,h\,{\left( 1 + m \right) }^2\,\lambda \,\left( -4 + {\left( 1 - m
    \right) }^2\,\lambda \right) + 4\,h^2\,\left( -8 + {\left( 1 - m
    \right) }^2\,\left( 1 + m \right) \,\lambda \right)  \\
    &\quad +
    {\left( 1 + m
    \right) }^3\,\lambda \,\left( -4 - 2\,\left( 1 + m \right) \,\lambda +
    {\left( 1 - m \right) }^2\,{\lambda }^2 \right)
   \\[1mm]
  \Delta_2 &=-\frac{\big(\overline r -4(1+m)\big)^2 u^2}{4rp^2\overline r ^2}
\end{align}
and $m$ is the solution of
\begin{align} \label{eqn:eqn_m_spins_psiplus2}
p \overline r (h+\lambda m - 1) \sqrt{r} = 
s u \,\big(\overline r - 4(1+m)\big)  +
\overline r\, \big(4hm-\lambda(1+m)(1-3m)\big)
\end{align}
such that $\psi^{(N)}(s)$ has the highest value.\\

The values of the first two  cumulants read
\begin{align}
 \frac{1}{N\,t} \langle Q_+ \rangle     &= 
   2\rho\ln 2 +\frac{1}{4 N} 
   \left[
          \lambda-\frac{1}{1-\rho}
          -\frac{8\lambda \rho (1-\rho)^2}
                {\big(1-\lambda(1-\rho)\big)^2}
           \ln 2
   \right]
   +{\cal O}(1/N^2) \\
 \frac{1}{N\,t} \langle Q_+^2 \rangle_c &= 
   \rho (\ln 2)^2 +\rho \left(\frac
      {1-2\rho+\lambda(1-\rho)^2}
      {1-\lambda(1-\rho)^2}\right)^2 \, (\ln 2)^2
   +{\cal O}(1/N)\\
\end{align}
The contact process also raises interest~\cite{deroulersmonasson} in related computationally motivated problems where similar absorbing-state phase transitions have been identified. We believe that not only the KS entropy, but also the pieces of information contained in $\tilde{P}_+$ or $\tilde{P}_K$, could shed a new light, with quantitative tools, on dynamical complexity issues. 

\section{Outlook}

Before concluding, we would like to discuss \cite{tailleur} on a simple example, namely
Brownian motion, the difference between the Markov approach we were dealing
with in this paper, and another possible approach which also generated
a lot of literature
in the field of dynamical and chaotic properties of systems.

Let us first adopt the Lorentz gas picture~\cite{dorfman} in which a particle is scattered by randomly placed obstacles. Over large distances, the particle is seen to perform a diffusive motion. Furthermore, two infinitesimally close-by particles will quickly follow exponentially diverging routes. This is a chaotic system. A Lorentz gas is well approximated by a Markov process. The possibility of choosing a variety of infinitesimally close initial conditions, leading to very different trajectories, is replaced with the drawing of random numbers whose net effect is to account for the chaotic nature of the Lorentz gas. Within this approach, such local characterization of chaos like individual Lyapunov exponents cannot be accessed.

An opposite approach to Brownian motion is the modeling in terms of a Langevin equation, say for the particle velocity, which evolves under the effect of an external position independent --~yet random~-- force. Within this picture~\cite{graham,gozzireuter,tanasenicolakurchan}, the random force is viewed as an external field. Two close-by initial conditions will be subjected to the same realization of the random force. Within this picture, a simple Brownian motion is not a chaotic system. What can possibly make it chaotic lies in space-dependent forces due to interactions or to an external field.

The difference in the two pictures lies in the observation scale compared with the intrinsic correlation length of the surrounding medium. In the first approach, the noise source is very short range correlated in space, but with long range time correlations. In the second approach, this is the exact opposite situation. When computing a Lyapunov exponent, before deciding which picture applies, one must compare the typical physical scales of the medium giving birth to a chaotic behavior. For times short with respect to the correlation time scale and distances large with respect to the correlation length, the first approach 
--~the Markov one~-- applies.

If this is the case,
we have shown that the thermodynamic formalism can successfully be
applied to Markov dynamics with continuous time, provided that the proper interpretation
is used for the definition of the dynamical partition function.
In particular, a finite \KS-entropy can be defined. This opens the door to explicit expressions for realistic systems.

Besides, we have embedded this formalism into a more general picture.
Indeed, the dynamical partition function can be expressed as the generating function
of an observable. By noticing that other observables could be used as well,
we are able to relate the quantities used in the thermodynamic formalism with those
involved in the much studied Lebowitz-Spohn~\cite{lebowitzspohn} fluctuation theorem. 
We also show on specific examples that the simplest observable one could think of,
namely the number $K$ of transitions occurring in a given time, is not as trivial as
one could think and contains already some relevant information on the system.
For example, for the infinite range Ising model, the cumulant generating function
of $K$ already indicates that a dynamical phase transition occurs in the low temperature
phase. This is confirmed by the calculation of the more sophisticated topological pressure.

We found also that one can gain some insight into these dynamical phase transitions
by looking at a new object: the aforementioned cumulant generating function
was obtained as the largest eigenvalue of a certain operator. If one also
computes the associated eigenvector, one can build a quantity that weights the trajectories
depending on  the value the observable takes along them. In the example of the
infinite range Ising model, this allows to show that the dynamical phase transition
which occurs below the critical temperature
gives rise to a splitting of the trajectories into two families,
respectively typical of a disordered and of an ordered phase.

The general unifying picture behind all this is that of a Gibbs ensemble construction carried out over the space of dynamical trajectories, rather than over microscopic states.\\

We have illustrated our approach on several physical examples (an interacting lattice gas, a system exhibiting an equilibrium second order phase transition and one with a nonequilibrium phase transition). Our setup has allowed us to provide an intrinsically dynamical picture to phenomena that are always interpreted in static terms.
This constitutes a powerful tool that longs to be applied to systems for which no static phenomena has ever been identified, like those possessing glassy dynamics. It is tempting to speculate that ageing and other dynamical features of glasses will be identified with a sharp signature on some appropriately chosen dynamical potentials like those considered throughout this work.
Some of these ideas can already be found in \cite{merollegarrahanchandler,jackgarrahanchandler}.
But before addressing these challenging issues, many questions remain to be answered for more conventional systems. As far as lattice gases are concerned, the general dependence of the KS entropy on the diffusion constant and the compressibility is one such question. Driving a lattice gas into a nonequilibrium steady-state (with a bulk or boundary field) leads to distinct dynamical features.
How do these reflect on the dynamical partition function?
In the vicinity of a second-order transition, the dynamics possesses universal features, so that the dynamical potentials $\psi_A(s)$ introduced in this paper will obey universal scaling laws. Which are these? May be some universal scaling functions as the one
found in \cite{derridaappert} could emerge.
The influence of quenched disorder, generically known to slow the dynamics down is one more open research route.\\

\noindent {\bf Acknowledgments:} The authors wish to thank T. Delattre for his participation in the early stage of this work, along with J. Tailleur, J. Kurchan and H. van Beijeren for their many helpful critical comments.
\appendix
\section{Non-hermitian quadratic operators and Bogoliubov-like transformation}
\label{appendixA}

Holstein-Primakoff expansions of our evolution operators $\mathbb{W}_A$ for infinite-range models often lead to a 
``Hamiltonian'' $\hat H$ that is quadratic in creation and annihilation operators $a$ and $a^\dag$
\begin{equation}
 \hat H = X a^2 + 2 Z a^\dag a + Y (a^\dag)^2
\end{equation}
We are interested in the lowest energy level of $\hat H$. In order that the latter exists we shall have to
 assume that $\Delta^2 = Z^2-XY > 0$ and $Y\leq 0$. 

Performing the similarity transformation $P_1^{-1} (\ldots) P_1$ with
\begin{equation}
 P_1 = e^{\frac{Z-\Delta}{2Y} a^2}
\end{equation}
does not alter $a$ while it shifts $a^\dag$ according to
\begin{equation} \label{eqn:P1}
 P_1^{-1} a^\dag P_1 = a^\dag - \frac{Z-\Delta}{Y}a
\end{equation}
Its purpose is to remove the $a^2$ term in $\hat H$:
\begin{equation}
 \hat H_1 = P_1^{-1} \hat H P_1 = Y (a^\dag)^2 + 2 \Delta a^\dag a +\Delta-Z
\end{equation}
We now introduce the operator
\begin{equation}
 P_2 = e^{-\frac{Y}{4\Delta} (a^\dag)^2}
\end{equation}
It commutes with $a^\dag$ and shifts $a$ according to
\begin{equation}\label{eqn:P2}
 P_2^{-1} a P_2 = a - \frac{Y}{2\Delta}a^\dag
\end{equation}
Acting on $\hat H_1$, it yields
\begin{equation}\label{eqn:H2}
 \hat H_2 = P_2^{-1} \hat H P_2 = 2 \Delta a^\dag a +\Delta-Z
\end{equation}
As the similarity transformations~(\ref{eqn:P1}) and~(\ref{eqn:P2}) do not
modify the spectrum of~$\hat H(s)$, we see that the lowest energy level
of~$\hat H(s)$ is $\Delta-Z$. When $H$ is Hermitian ($X=Y$), the Bogoliubov
transformation leads to exactly the same result. However, when $H$ is not
Hermitian, the Bogoliubov transformation cannot be implemented: contrary
to~(\ref{eqn:P1}) and~(\ref{eqn:P2}), it does not transform~$a$ and~$a^\dag$
independently, which was required here to obtain~\eqref{eqn:H2}.

\end{document}